\documentclass[journal,draftcls,onecolumn,12pt,twoside]{IEEEtran}
\usepackage{graphicx}
\usepackage{amsmath, amsfonts, amssymb, amsthm}
\usepackage{enumerate}
\usepackage{multirow}
\usepackage{color}
\usepackage{longtable}
\theoremstyle{definition} \newtheorem{theorem}{Theorem}
\theoremstyle{definition} 
\theoremstyle{definition} \newtheorem{proposition}[theorem]{Proposition}
\theoremstyle{definition} 
\theoremstyle{definition} \newtheorem{lemma}[theorem]{Lemma}
\theoremstyle{definition} \newtheorem{algorithm}[theorem]{Algorithm}

\theoremstyle{definition} 
\theoremstyle{definition} 
\theoremstyle{definition} \newtheorem*{example}{Example}
\theoremstyle{definition} 
{\end{list}}

\allowdisplaybreaks

\begin{document}
%
\title{Circular-Shift Linear Network Codes \\ with Arbitrary Odd Block Lengths}
\author{\IEEEauthorblockN{Qifu~Tyler~Sun\textsuperscript{\dag},~Hanqi~Tang\textsuperscript{\dag},~Zongpeng~Li\textsuperscript{\ddag},~Xiaolong Yang\textsuperscript{\dag},~and~Keping~Long\textsuperscript{\dag}}
\IEEEauthorblockA{\textsuperscript{\dag}Department of Communication Engineering, University of Science and Technology Beijing, China \\
\textsuperscript{\ddag} School of Computer Science, Wuhan University, China\\
\thanks{Q. T. Sun (qfsun@ustb.edu.cn) is the corresponding author.} 
}
}
\maketitle
\sloppy

\begin{abstract}
Circular-shift linear network coding (LNC) is a class of vector LNC with low encoding and decoding complexities, and with local encoding kernels chosen from cyclic permutation matrices. When $L$ is a prime with primitive root $2$, it was recently shown that a scalar linear solution over GF($2^{L-1}$) induces an $L$-dimensional circular-shift linear solution at rate $(L-1)/L$. In this work, we prove that for arbitrary odd $L$, every scalar linear solution over GF($2^{m_L}$), where $m_L$ refers to the multiplicative order of $2$ modulo $L$, can induce an $L$-dimensional circular-shift linear solution at a certain rate. Based on the generalized connection, we further prove that for such $L$ with $m_L$ beyond a threshold, every multicast network has an $L$-dimensional circular-shift linear solution at rate $\phi(L)/L$, where $\phi(L)$ is the Euler's totient function of $L$. An efficient algorithm for constructing such a solution is designed. Finally, we prove that every multicast network is asymptotically circular-shift linearly solvable.
\end{abstract}
\sloppy

\section{Introduction}
A multicast network is a finite directed acyclic multigraph, with a unique source node $s$ and a set $T$ of receivers. In a multicast network, the source $s$ generates $\omega$ binary sequences of length $L'$, and every edge transmits a binary sequence of length $L$. A linear network coding (LNC) scheme computes an outgoing binary sequence from a non-source node $v$ as a linear function of the incoming binary sequences to $v$. It qualifies as an $L$-dimensional linear solution at rate $L'/L$ if every receiver can recover the $\omega$ source binary sequences of length $L'$ from its incoming binary sequences of length $L$.

On a binary sequence, a circular-shift operation implemented in software incurs negligible computational complexity, compared with bit-wise additions; circular-shift is also amenable to implementation through atomic hardware operations. In order to reduce the encoding complexity of LNC, Ref. \cite{Xiao07}-\cite{Tang_Sun_Circular-shift_LNC} studied LNC schemes with circular-shifts as the linear operations on a binary sequence. Specifically, when $L$ is a large enough prime minus $1$, a low-complexity linear solution at rate $1$ was designed in \cite{Xiao07} for a special class of multicast networks known as Combination Networks. %
The LNC schemes studied in \cite{Khojastepour10} are called {\em rotation-and-add} linear codes, and are applicable to an arbitrary multicast network. The ones studied in \cite{HouShum16} are called BASIC functional-repair regenerating codes. BASIC codes are discussed in the context of a distributed storage system, which is essentially equivalent to a multicast network. %
When $L > |T|$ is a prime with primitive root $2$, \emph{i.e.}, the multiplicative order of $2$ modulo $L$ is $L-1$, the existence of an $L$-dimensional rotation-and-add linear solution at rate $(L-1)/L$ and an $L$-dimensional BASIC functional-repair regenerating code at rate $(L-1)/L$ have been respectively shown in \cite{Khojastepour10} and \cite{HouShum16}, through the approach of cyclic convolutional coding.

More recently, circular-shift LNC was formulated in \cite{Tang_Sun_Circular-shift_LNC} in the context of a general acyclic network and from the perspective of \emph{vector} LNC. %
Compared with the conventional scalar LNC approach (See, e.g., \cite{LiYeungCai03}\cite{KoetterMedard03}), which models binary sequences as elements in GF($2^L$), vector LNC (See, e.g., \cite{Medard03}-\cite{Etzion18}) models binary sequences as vectors in GF($2$)$^L$. The coding operations performed at intermediate nodes by scalar LNC and by vector LNC are linear functions over GF($2^L$) and over the ring of $L \times L$ binary matrices, respectively. 
Under the framework of vector LNC, the linear coding operation of circular-shifts on a binary sequence $[m_1~m_2~\ldots~m_L]$ by $1 \leq j \leq L-1$ positions to the right can simply be expressed as %
$
[m_1~  m_2~ \ldots~ m_L]\mathbf{C}_L^j
=[m_{L-j+1}~\ldots~m_L~ m_1~ \ldots~ m_{L-j}],
$
where $\mathbf{C}_L$ refers to the following $L\times L$ cyclic permutation matrix over GF($2$)
\begin{equation}
\label{eqn:cyclic_permutation_matrix}
\setlength{\arraycolsep}{3.0pt}
\renewcommand{\arraystretch}{0.7}
\mathbf{C}_L = \left[\begin{matrix}
0 & 1 & 0 & \ldots & 0 \\
0 & 0 & 1 & \ddots & 0 \\
0 & \ddots & \ddots & \ddots & 0 \\
0 & \ddots & \ddots & 0 & 1 \\
1 & 0 & \ldots & 0 & 0 \end{matrix}\right].
\end{equation}
A key advantage of such formulation utilized in \cite{Tang_Sun_Circular-shift_LNC} is that when $L$ is odd, the cyclic permutation matrix $\mathbf{C}_L^j$ can be diagonalized in a way for easier manipulation. Prior to \cite{Tang_Sun_Circular-shift_LNC}, such a diagonalization manipulation on $\mathbf{C}_L$ has also been adopted in the rank analysis of quasi-cyclic LDPC codes \cite{HuangQin_TCom_10}\cite{HuangQin_TIT_12} as well as certain quasi-cyclic stabilizer quantum LDPC codes \cite{XieShane_TCom_18}. %

When $L$ is a prime with primitive $2$, it was revealed in \cite{Tang_Sun_Circular-shift_LNC} that every scalar linear solution over GF($2^{L-1}$) induces an $L$-dimensional circular-shift linear solution at rate $(L-1)/L$. Thus, a rotation-and-add linear solution considered in \cite{Khojastepour10} and a BASIC functional-repair regenerating code considered in \cite{HouShum16} can be efficiently constructed via the efficient construction of a scalar linear solution.

In order to make the design of circular-shift LNC more flexible, in the present paper, we continue to investigate an intrinsic connection between scalar LNC and circular-shift LNC for an \emph{arbitrary odd block length} $L$, in the context of multicast networks. %
For multicast networks, the work in \cite{Tang_Sun_Circular-shift_LNC}, which considers prime $L$ with primitive root $2$, is a special case of the present work. Under such an assumption on $L$, one of the technical keys that make the analysis relatively easier is that the polynomial $1+x+\ldots+x^{L-1}$ is irreducible over GF($2$). For general odd $L$, as $1+x+\ldots+x^{L-1}$ is no longer irreducible, we need to further deal with its structure to obtain the more general framework between scalar LNC and circular-shift LNC. Though a similar approach to \cite{Tang_Sun_Circular-shift_LNC} can be adopted to theoretically obtain a circular-shift linear solution from a scalar linear solution, the rate of the induced circular-shift linear solution is not necessarily $(L-1)/L$. Moreover, the design of a concomitant source encoding matrix, which transforms the binary sequences of length $L'$ generated at the source $s$ to binary sequences of lengths $L$ transmitted along outgoing edges of $s$, becomes more challenging. Therefore, it deserves our further investigation in this paper. %
The main contributions and the organization of this paper are summarized as follows:
\begin{itemize}
\item After reviewing preliminary literature of LNC in Section II, we introduce a method in Section III to obtain an $L$-dimensional circular-shift linear code from an arbitrary scalar linear code over GF($2^{m_L}$), where $m_L$ refers to the multiplicative order of $2$ modulo $L$. Based on a rank analysis between the scalar linear code and the induced circular-shift linear code, we further turn the circular-shift linear code into a circular-shift linear solution at a certain rate $L'/L$ by explicitly constructing an $\omega L' \times \omega L$ source encoding matrix. %
\item Under the general framework, in Section IV, we first prove the existence of an $L$-dimensional circular-shift linear solution at rate $\phi(L)/L$, as long as $m_L$ is large enough. Here $\phi(L)$ refers to the Euler's totient function of $L$ and one of the specific sufficient bounds on $m_L$ is the number of receivers. An efficient algorithm to construct such a solution, via a flow path approach, is also proposed. %
\item Stemming from the existence of a circular-shift linear solution at rate $\phi(L)/L$, in Section V, we provide a positive answer to an open conjecture in \cite{Tang_Sun_Circular-shift_LNC}: every multicast network is asymptotically circular-shift linearly solvable.
\end{itemize}

In addition to the detailed proof of lemmas, theorems, and propositions, frequently used notation is also listed in the Appendix for reference.


\section{Preliminaries}
In the present paper, we consider a multicast network, which is modeled as a finite directed acyclic multigraph, with a unique source node $s$ and a set $T$ of receivers. %
For a node $v$ in the network, denote by $\mathrm{In}(v)$ and $\mathrm{Out}(v)$, respectively, the set of its incoming and outgoing edges. A pair $(d,e)$ of edges is called an adjacent pair if there is a node $v$ with $d \in \mathrm{In}(v)$ and $e \in \mathrm{Out}(v)$. Every edge $e$ has unit capacity, that is, it transmits one data unit, which is an $L$-dimensional row vector $\mathbf{m}_e$ of binary data symbols, per edge use. %
For every receiver $t\in T$, based on the $|\mathrm{In}(t)|$ received data units, the goal is to recover the $\omega$ source data units generated by $s$. The maximum flow from $s$ to $t$, which is equal to the number of edge-disjoint paths from $s$ to $t$, is assumed to be $\omega$. Without loss of generality, assume $|\mathrm{Out}(s)| = |\mathrm{In}(t)| = \omega$, and there is not any edge leading from $s$ to $t$. A topological order is also assumed on $E$ led by edges in $\mathrm{Out}(s)$.

An \emph{$L$-dimensional vector linear code} $(\mathbf{K}_{d,e})$ (over GF(2) and at rate $1$) is an assignment of a \emph{local encoding kernel} $\mathbf{K}_{d,e}$, which is an $L\times L$ matrix over GF(2), to every pair $(d, e)$ of edges such that $\mathbf{K}_{d,e}$ is the zero matrix $\mathbf{0}$ when $(d, e)$ is not an adjacent pair. For every edge $e$ emanating from a non-source node $v$, the data unit vector $\mathbf{m}_e = \sum_{d\in \mathrm{In}(v)} \mathbf{m}_d\mathbf{K}_{d,e}$. %
Every vector linear code uniquely determines a global encoding kernel $\mathbf{F}_{e}$, which is an $\omega L\times L$ matrix over GF(2), for every edge $e$. %
A vector linear code is a \emph{vector linear solution} if for every receiver $t \in T$, the column-wise juxtaposition%
\footnote{Throughout this paper, the notion $[\mathbf{A}_e]_{e\in E'}$ will always refer to \emph{column-wise} juxtaposition of matrices $\mathbf{A}_e$ with $e$ orderly chosen from a subset $E'$ of $E$, where $\mathbf{A}_e$ may degenerate to vectors.} $[\mathbf{F}_e]_{e\in \mathrm{In}(t)}$ has full rank $\omega L$. %
A $1$-dimensional vector linear code is a scalar linear code, in which case we shall use the scalar symbol $k_{d,e}$ and the vector symbol $\mathbf{f}_e$ to denote the local and global encoding kernels, respectively.

As formulated in \cite{Tang_Sun_Circular-shift_LNC}, an \emph{$L$-dimensional circular-shift linear code of degree $\delta$}, $0 \leq \delta \leq L$,  is an $L$-dimensional vector linear code with local encoding kernels selected from \begin{equation}
\label{eqn:set_C_delta}
\mathcal{C}_\delta = \left\{\sum\nolimits_{j=0}^{L-1}a_j\mathbf{C}_L^j: a_j \in \{0, 1\} \subset \mathbb{Z}, \sum\nolimits_{j = 0}^{L-1} a_j \leq \delta \right\},
\end{equation}
that is, from matrices that can be written as summation of at most $\delta$ cyclic permutation matrices. There exist multicast networks that do not have an $L$-dimensional circular-shift linear solution of degree $\delta$ for any $L$ and $\delta$ \cite{Tang_Sun_Circular-shift_LNC}. However, when $L$ is a prime with primitive root $2$, an $L$-dimensional circular-shift linear solution at rate $(L-1)/L$ can be readily obtained from a scalar linear solution over GF($2^{L-1}$) subject to some local encoding kernel constraints, where an $L$-dimensional (fractional) linear code at rate $L'/L$ is a variation of an $L$-dimensional vector linear code with the following differences (See, e.g., \cite{Connelly_Zeger17}\cite{Tang_Sun_Circular-shift_LNC}):
the $\omega$ data units $\mathbf{m}'_1, \ldots, \mathbf{m}'_\omega$ generated at $s$ are $L'$-dimensional row vectors over GF(2), and each of the $L$ binary data symbols in $\mathbf{m}_e$, $e \in \mathrm{Out}(s)$, is a GF(2)-linear combination of the ones in $\mathbf{m}'_1, \ldots, \mathbf{m}'_\omega$, \emph{i.e.},
\[
[\mathbf{m}_e]_{e \in \mathrm{Out}(s)} = [\mathbf{m}_i']_{1 \leq i \leq \omega}\mathbf{G}_s
\]
for some $\omega L' \times \omega L$ source encoding matrix $\mathbf{G}_s$ over GF(2). %

For brevity, an $L$-dimensional linear code at rate $L'/L$ will be called an $(L', L)$ linear code. %
An $(L', L)$ linear code qualifies as a linear solution if for every receiver $t \in T$, the $\omega L' \times \omega L$ matrix $\mathbf{G}_s[\mathbf{F}_e]_{e\in \mathrm{In}(t)}$ has full rank $\omega L'$. %
For an $(L', L)$ linear solution, each receiver $t$ has an $\omega L \times \omega L'$ decoding matrix $\mathbf{D}_t$ over GF(2) such that
\[
\mathbf{G}_s[\mathbf{F}_e]_{e\in \mathrm{In}(t)}\mathbf{D}_t = \mathbf{I}_{\omega L'},
\]
where $\mathbf{I}_{\omega L'}$ refers to the identity matrix of size $\omega L'$. Based on $\mathbf{D}_t$, the $\omega$ source data units can be recovered at $t$ via
\[
[\mathbf{m}_e]_{e\in \mathrm{In}(t)}\mathbf{D}_t = \left([\mathbf{m}_i']_{1 \leq i \leq \omega}\mathbf{G}_s[\mathbf{F}_e]_{e\in \mathrm{In}(t)}\right)\mathbf{D}_t = [\mathbf{m}_i']_{1 \leq i \leq \omega}.
\]

As remarked in the previous section, a key reason for formulating circular-shift LNC from the perspective of vector LNC \cite{Tang_Sun_Circular-shift_LNC} is to exploit the following diagonalization of cyclic permutation matrices for odd $L$:
\begin{equation}
\label{eqn:cyclic_matrix_decomposition}
\mathbf{C}_L^j = \mathbf{V}_L \cdot \mathbf{\Lambda}_\alpha^j \cdot \mathbf{V}_L^{-1} ~~~~~ \forall j \geq 0
\end{equation}
where $\alpha$ is a primitive $L^{th}$ root of unity over GF($2$), $\mathbf{V}_L$ is the $L\times L$ Vandermonde matrix generated by $1, \alpha, \alpha^2, \ldots, \alpha^{L-1}$ over GF(2)($\alpha$), the minimal field containing GF($2$) and $\alpha$, and $\mathbf{\Lambda}_\alpha$ is the $L \times L$ diagonal matrix with diagonal entries $1, \alpha, \ldots, \alpha^{L-1}$.
Specifically,
\begin{equation}
\label{eqn:Vandermonde_p}
\mathbf{V}_L = \setlength{\arraycolsep}{3.0pt}
\renewcommand{\arraystretch}{0.7}
\left[
\begin{matrix}
1 & 1 & 1 & \ldots & 1 \\
1 & \alpha & \alpha^2 & \ldots & \alpha^{L-1} \\
\vdots & \vdots & \vdots & \ldots & \vdots \\
1 & \alpha^{L-1} & \alpha^{(L-1)2} &\ldots & \alpha^{(L-1)(L-1)}
\end{matrix}
\right],
\end{equation}
\begin{equation}
\mathbf{\Lambda}_\alpha =
\setlength{\arraycolsep}{3.0pt}
\renewcommand{\arraystretch}{0.7}
\left[\begin{matrix}
1 & 0 & \ldots & 0 \\
0 & \alpha & \ddots & \vdots \\
\vdots & \ddots & \ddots & 0 \\
0 & \ldots & 0 & \alpha^{L-1}
\end{matrix}\right],
\end{equation}
\begin{equation}
\label{eqn:Vandermond_L_inverse}
\mathbf{V}_L^{-1} =
\setlength{\arraycolsep}{3.0pt}
\renewcommand{\arraystretch}{0.7}
\left[
\begin{matrix}
1 & 1 & 1 & \ldots & 1 \\
1 & \alpha^{-1} & \alpha^{-2} & \ldots & \alpha^{-(L-1)} \\
\vdots & \vdots & \vdots & \ldots & \vdots \\
1 & \alpha^{-(L-1)} & \alpha^{-(L-1)2} & \ldots & \alpha^{-(L-1)(L-1)}
\end{matrix}
\right].
\end{equation}
Eq. (\ref{eqn:cyclic_matrix_decomposition}) will also facilitate us to establish a more general connection between circular-shift LNC and scalar LNC in this work.

One may refer to Appendix-\ref{appendix:List_of_Notation} for a list of frequently used notations in the present paper.


\section{Circular-shift LNC over Odd Block Lengths}
\subsection{General Framework}
Hereafter in this paper, let $L$ denote a positive odd integer, $m_L$ denote the multiplicative order of $2$ modulo $L$, and $\alpha$ be a primitive $L^{th}$ root of unity over GF($2$). Then, the minimum field containing both GF($2$) and $\alpha$ is GF($2^{m_L}$). When a scalar linear code over GF($2^{m_L}$) is denoted by $(k_{d,e}(\alpha))$, it means that every local encoding kernel $k_{d,e}(\alpha)$  is the evaluation of a defined polynomial $k_{d,e}(x)$ over GF($2$) by setting $x$ equal to $\alpha$.

When $L$ is a prime with primitive root $2$, $m_L = L-1$. In this special case, it has been revealed that in a general acyclic (multi-source multicast) network, every scalar linear solution over GF($2^{L-1}$) induces an $(L-1, L)$ circular-shift linear solution in a rather straightforward manner \cite{Tang_Sun_Circular-shift_LNC}. Actually, we next demonstrate that such construction of a circular-shift linear code also applies to the case that $L$ is an odd integer.

On a multicast network, consider a scalar linear code $(k_{d,e}(\alpha))$ over GF($2^{m_L}$), and let $(k_{d,e}(x))$ denote a corresponding defined scalar linear code $(k_{d,e}(x))$ over the polynomial ring GF($2$)$[x]$ such that $k_{d,e}(\alpha)$ is the evaluation of $k_{d,e}(x)$ by setting $x = \alpha$. For every $0 \leq j \leq L-1$, $(k_{d,e}(\alpha^j))$ also forms a scalar linear code over GF($2^{m_L}$), where $k_{d,e}(\alpha^j)$ is the evaluation of $k_{d,e}(x)$ by setting $x = \alpha^j$. Define an $L$-dimensional circular-shift linear code $(\mathbf{K}_{d,e})$ by
\begin{equation}
\label{eqn:scalar-to-vector-circular-shift}
\mathbf{K}_{d,e} = \left\{ \begin{matrix} \mathbf{0} & \quad \mathrm{if}~k_{d,e}(x) = 0 \\ k_{d,e}(\mathbf{C}_L) & \quad \mathrm{otherwise} \end{matrix} \right.
\end{equation}
where $k_{d,e}(\mathbf{C}_L)$ means an $L\times L$ matrix obtained via replacing $x$ by $\mathbf{C}_L$ in the polynomial $k_{d,e}(x)$. For an edge $e$, let $\mathbf{f}_e(x)$ denote its global encoding kernel determined by $(k_{d,e}(x))$, which is an $\omega$-dimensional vector defined over GF($2$)$[x]$. Thus, the global encoding kernel for edge $e$ determined by $(k_{d,e}(\alpha^j))$ and by $(\mathbf{K}_{d,e})$ can be respectively expressed as $\mathbf{f}_e(\alpha^j)$ and $\mathbf{F}_e = \mathbf{f}_e(\mathbf{C}_L)$. %

\begin{theorem}
\label{thm:rank-relation}
For every receiver $t$,
\begin{equation}
\label{eqn:rank_equation}
\mathrm{rank}\left([\mathbf{F}_e]_{e\in \mathrm{In}(t)}\right) = \sum\nolimits_{j = 0}^{L-1}\mathrm{rank}\left([\mathbf{f}_e(\alpha^j)]_{e\in \mathrm{In}(t)}\right).
\end{equation}
\begin{proof}
Please refer to Appendix-\ref{appendix:proof-rank-relation}.
\end{proof}
\end{theorem}

\begin{figure}[!t]
\centering
\scalebox{0.63}
{\includegraphics{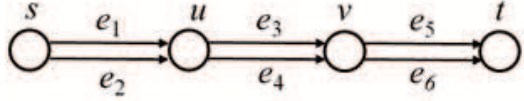}}
\caption{A network consists of four nodes.}
\label{Fig:4-node-network}
\end{figure}

\begin{example}
Consider the network depicted in Fig.\ref{Fig:4-node-network}, which consists of the source node $s$, two relay nodes and the receiver $t$. Assume $L = 9$. In this case, $m_L = 6$ and $\alpha$ is a root of $x^6 + x^3 + 1$, which divides $x^9+1$. Consider the following scalar linear code $(k_{d,e}(\alpha))$ over GF($2^6$):
\begin{equation*}
\begin{matrix}
k_{e_1,e_3}(\alpha) = k_{e_1,e_4}(\alpha) = k_{e_3,e_5}(\alpha) = 1 & k_{e_2,e_4}(\alpha) = 1+\alpha^3 \\
k_{e_2,e_3}(\alpha) = k_{e_4,e_5}(\alpha) = k_{e_3,e_6}(\alpha) = 0 & k_{e_4,e_6}(\alpha) = 1+\alpha^6 \\
\end{matrix}
\end{equation*}
Determined by $(k_{d,e}(\alpha))$, the global encoding kernels for incoming edges to $t$ are
\[
[\mathbf{f}_e(\alpha)]_{e\in \mathrm{In}(t)} = 
\left[\begin{matrix} 1 & 1+\alpha^6 \\ 0 & (1+\alpha^3)(1+\alpha^6) \end{matrix} \right] =
\left[\begin{matrix} 1 & 1+\alpha^6 \\ 0 & 1 \end{matrix} \right].
\]

When $k_{e_2,e_4}(\alpha) = 1+\alpha^3$ and $k_{e_4,e_6}(\alpha) = 1+\alpha^6$ are respectively regarded as the evaluation of defined polynomials $k_{e_2,e_4}(x) = 1+x^3$ and $k_{e_4,e_6}(x) = 1+x^6$, $k_{e_2,e_4}(\alpha^3) = 1+\alpha^9 = 0$ and $k_{e_4,e_6}(\alpha^3) = 1+\alpha^{18} = 0$. Thus, determined by $(k_{d,e}(\alpha^3))$, the global encoding kernels for incoming edges to $t$ are
$[\mathbf{f}_e(\alpha^3)]_{e\in \mathrm{In}(t)} =
\setlength{\arraycolsep}{3.0pt}
\renewcommand{\arraystretch}{0.7}
\left[\begin{matrix} 1 & 0 \\ 0 & 0 \end{matrix} \right]$. Note that $\mathbf{f}_{e_6}(\alpha^3) = [0~0]^T$ is calculated based on $k_{d,e}(\alpha^3)$ and it cannot be obtained from $\mathbf{f}_{e_6}(\alpha) = [1+\alpha^6~1]^T$ by simply replacing $\alpha$ with $\alpha^3$. %
One can further verify that
$[\mathbf{f}_e(\alpha^j)]_{e\in \mathrm{In}(t)} = \setlength{\arraycolsep}{3.0pt}
\renewcommand{\arraystretch}{0.7}
\left[\begin{matrix} 1 & 0 \\ 0 & 0 \end{matrix} \right]$
when $j \in \{0, 3, 6\}$, %
$[\mathbf{f}_e(\alpha^j)]_{e\in \mathrm{In}(t)} =
\setlength{\arraycolsep}{3.0pt}
\renewcommand{\arraystretch}{0.7}
\left[\begin{matrix} 1 & 1+\alpha^6 \\ 0 & 1 \end{matrix} \right]$ when $j \in \{1, 4, 7\}$, %
and $[\mathbf{f}_e(\alpha^j)]_{e\in \mathrm{In}(t)} = \setlength{\arraycolsep}{3.0pt}
\renewcommand{\arraystretch}{0.7}
\left[\begin{matrix} 1 & 1+\alpha^3 \\ 0 & 1 \end{matrix} \right]$ when $j \in \{2, 8, 5\}$. Thus, $\sum\nolimits_{j = 0}^{L-1}\mathrm{rank}\left([\mathbf{f}_e(\alpha^j)]_{e\in \mathrm{In}(t)}\right) = 15$.

Now consider the circular-shift linear code $(\mathbf{K}_{d,e})$ induced from $(k_{d,e}(\alpha))$ via (\ref{eqn:scalar-to-vector-circular-shift}). Determined by it,
\[
[\mathbf{F}_e]_{e\in \mathrm{In}(t)} = \setlength{\arraycolsep}{3.0pt}
\renewcommand{\arraystretch}{0.7}
\left[\begin{matrix} \mathbf{I}_9 & \mathbf{I}_9+\mathbf{C}_9^6 \\ \mathbf{0} & (\mathbf{I}_9+\mathbf{C}_9^3)(\mathbf{I}_9+\mathbf{C}_9^6) \end{matrix} \right],
\]
the rank of which is also $15$.  %

Next, as $\alpha^6 + \alpha^3 + 1 = 0$, $k_{e_2,e_4}(\alpha) = 1+\alpha^3$ and $k_{e_4,e_6}(\alpha) = 1+\alpha^6$ can be respectively expressed as $\alpha^6$ and $\alpha^3$ and regarded as the evaluation of defined polynomials $k_{e_2,e_4}(x) = x^6$ and $k_{e_4,e_6}(x) = x^3$. Under this new setting,
$[\mathbf{f}_e(\alpha^j)]_{e\in \mathrm{In}(t)} =
\setlength{\arraycolsep}{3.0pt}
\renewcommand{\arraystretch}{0.7}
\left[\begin{matrix} 1 & 1 \\ 0 & 1 \end{matrix} \right]$ when $j \in \{0, 3, 6\}$, %
$[\mathbf{f}_e(\alpha^j)]_{e\in \mathrm{In}(t)} =
\setlength{\arraycolsep}{3.0pt}
\renewcommand{\arraystretch}{0.7}
\left[\begin{matrix} 1 & \alpha^3 \\ 0 & 1 \end{matrix} \right]$ when $j \in \{1, 4, 7\}$, %
and $[\mathbf{f}_e(\alpha^j)]_{e\in \mathrm{In}(t)} = \setlength{\arraycolsep}{3.0pt}
\renewcommand{\arraystretch}{0.7}
\left[\begin{matrix} 1 & 1+\alpha^6 \\ 0 & 1 \end{matrix} \right]$ when $j \in \{2, 8, 5\}$. Thus, $\sum\nolimits_{j = 0}^{L-1}\mathrm{rank}\left([\mathbf{f}_e(\alpha^j)]_{e\in \mathrm{In}(t)}\right) = 18$. %
For the corresponding circular-shift linear code $(\mathbf{K}_{d,e})$ defined by (\ref{eqn:scalar-to-vector-circular-shift}),
$[\mathbf{F}_e]_{e\in \mathrm{In}(t)} = \setlength{\arraycolsep}{3.0pt}
\renewcommand{\arraystretch}{0.7}
\left[\begin{matrix} \mathbf{I}_9 & \mathbf{C}_9^3 \\ \mathbf{0} & \mathbf{I}_9 \end{matrix} \right]$, the rank of which equals $18$ too.  \hfill $\blacksquare$%
\end{example}

Compared with the results in \cite{Tang_Sun_Circular-shift_LNC}, Theorem \ref{thm:rank-relation} establishes a more fundamental connection between circular-shift LNC and scalar LNC, which not only holds for an arbitrary odd block length $L$, but also for an arbitrary scalar linear code $(k_{d,e}(\alpha))$ over GF($2^{m_L}$). %
On one hand, it justifies that in the application of circular-shift LNC, the 1-bit redundancy during transmission is inevitable in the following sense. In order to make $[\mathbf{F}_e]_{e\in \mathrm{In}(t)}$ full rank $\omega L$ for an $L$-dimensional circular-shift linear code, according to Eq. (\ref{eqn:rank_equation}), the $\omega\times \omega$ matrix $[\mathbf{f}_e(1)]_{e\in \mathrm{In}(t)}$ determined by the scalar linear code $(k_{d,e}(1))$ needs to be full rank $\omega$. Since $(k_{d,e}(1))$ is defined over GF($2$), it directly endows low implementation complexity and there is no need to consider LNC at all.

On the other hand, it asserts that every scalar linear solution is possible to induce an $(L', L)$ circular-shift linear solution at a certain rate $L'/L$.

For instance, as proved in \cite{Tang_Sun_Circular-shift_LNC}, when $L$ is a prime with primitive root $2$, if an arbitrary scalar linear code $(k_{d,e}(\alpha))$ over GF($2^{L-1}$) qualifies as a linear solution, then the scalar linear code $(k_{d,e}(\alpha^j))$ is a linear solution for all $1 \leq j \leq L-1$ too. This property of $(k_{d,e}(\alpha))$, together with Theorem \ref{thm:rank-relation}, guarantees that the $L$-dimensional circular-shift linear code $(\mathbf{K}_{d,e})$ defined by (\ref{eqn:scalar-to-vector-circular-shift}) satisfies $\mathrm{rank}([\mathbf{F}_e]_{e \in \mathrm{In}(t)}) \geq \omega(L-1)$ for every receiver $t$. Thus, after appropriately designing a source encoding matrix $\mathbf{G}_s$, we can obtain an $(L-1,L)$ circular-shift linear solution $(\mathbf{K}_{d,e})$ induced from $(k_{d,e}(\alpha))$.


Stemming from this idea, we next deal with the case that the block length $L$ is an arbitrary odd integer, so that the circular-shift linear code $(\mathbf{K}_{d,e})$ constructed from a scalar linear code $(k_{d,e}(\alpha))$ by (\ref{eqn:scalar-to-vector-circular-shift}) can constitute a linear solution at a certain rate via embedding an appropriate source encoding matrix $\mathbf{G}_s$. First we observe the following property on a scalar linear code $(k_{d,e}(\alpha))$ over GF($2^{m_L}$).

\begin{lemma}
\label{lemma:scalar_linear_solution_induce_another_scalar}
If $(k_{d,e}(\alpha))$ is a scalar linear solution, then for every $j \geq 0$, the scalar linear code $(k_{d,e}(\alpha^{2^j}))$ qualifies as a linear solution too.
\begin{proof}
Consider a receiver $t$ and a nonnegative integer $j$. It can be shown that the mapping $\sigma_j: \mathrm{GF}(2^{m_L}) \rightarrow \mathrm{GF}(2^{m_L})$ defined by $\sigma_j(\beta) = \beta^{2^j}$ is an automorphism of GF($2^{m_L}$) that fixes the elements in GF($2$) (See, e.g., Theorem 2.21 in \cite{Lidl_Finite_Fields}). Thus, the full rank of $[\mathbf{f}_e(\alpha^{2^j})]_{e \in \mathrm{In}(t)}$ can be readily implied by the full rank of $[\mathbf{f}_e(\alpha)]_{e \in \mathrm{In}(t)}$ since  $\mathrm{det}\left([\mathbf{f}_e(\alpha^{2^j})]_{e \in \mathrm{In}(t)}\right) = \mathrm{det}\left([\mathbf{f}_e(\alpha)]_{e \in \mathrm{In}(t)}\right)^{2^j} \neq 0$.
\end{proof}
\end{lemma}

Let $\mathcal{J}$ be the set of integers between $0$ and $L-1$ such that the scalar linear code $(k_{d,e}(\alpha^j))$ over GF($2^{m_L}$) is a linear solution. As a consequence of Lemma \ref{lemma:scalar_linear_solution_induce_another_scalar}, $\mathcal{J}$ is closed under multiplication by $2$ (modulo $L$). Let $J = |\mathcal{J}|$. %
Denote by $\tilde{\mathbf{I}}_\mathcal{J}$ the $J\times L$ matrix obtained from $\mathbf{I}_L$ by deleting the $(j+1)^{st}$ row whenever $j \notin \mathcal{J}$, and by $\tilde{\mathbf{V}}$ the $J\times J$ matrix obtained from $\tilde{\mathbf{I}}_\mathcal{J}\mathbf{V}_L$ by restricting to the first $J$ columns. For instance, when $L = 15$ and $\mathcal{J} = \{1, 2, 4, 8\}$,
\begin{equation}
\label{eqn:V_tilde_example}
\tilde{\mathbf{I}}_\mathcal{J}\mathbf{V}_L = \setlength{\arraycolsep}{2.0pt}
\renewcommand{\arraystretch}{0.7}
\left[
\begin{matrix}
1 & \alpha & \alpha^2 & \alpha^3 & \alpha^4 & \ldots & \alpha^{13} & \alpha^{14} \\
1 & \alpha^2 & \alpha^4 & \alpha^6 & \alpha^8 & \ldots & \alpha^{11} & \alpha^{13} \\
1 & \alpha^4 & \alpha^8 & \alpha^{12} & \alpha & \ldots & \alpha^{7} & \alpha^{11} \\
1 & \alpha^8 & \alpha & \alpha^9 & \alpha^2 & \ldots & \alpha^{14} & \alpha^7
\end{matrix}
\right],~
\tilde{\mathbf{V}} = \setlength{\arraycolsep}{2.0pt}
\renewcommand{\arraystretch}{0.7}
\left[
\begin{matrix}
1 & \alpha & \alpha^2 & \alpha^3 \\
1 & \alpha^2 & \alpha^4 &\alpha^6 \\
1 & \alpha^4 & \alpha^8 & \alpha^{12}\\
1 & \alpha^8 & \alpha & \alpha^9
\end{matrix}
\right],
\end{equation}
where $\alpha \in \mathrm{GF}(2^4)$ is a primitive $15^{th}$ root of unity. %
As $\tilde{\mathbf{V}}$ can be regarded as a $J\times J$ Vandermonde matrix generated by $\alpha^j, j\in \mathcal{J}$, it is invertible. Define $\mathbf{G}$ and $\mathbf{G}_s$, respectively, to be the following $J \times L$ and $J\omega \times L\omega$ matrix
\begin{equation}
\label{eqn:G_definition}
\mathbf{G} = \tilde{\mathbf{V}}^{-1}\tilde{\mathbf{I}}_\mathcal{J}\mathbf{V}_L^{-1},~
\mathbf{G}_s = \mathbf{I}_\omega\otimes\mathbf{G},
\end{equation}
where $\otimes$ denotes the Kronecker product.

\begin{lemma}
\label{lemma:G_over_GF2}
Every entry in $\mathbf{G}$, and hence in $\mathbf{G}_s$, belongs to GF($2$).
\begin{proof}
Please refer to Appendix-\ref{appendix:G_over_GF2}.
\end{proof}
\end{lemma}

As an example, when $L = 15$ and $\mathcal{J} = \{1, 2, 4, 8\}$, $\tilde{\mathbf{V}}$ is given in (\ref{eqn:V_tilde_example}), and thus
\setcounter{MaxMatrixCols}{15}
\begin{align*}
\mathbf{G} = \tilde{\mathbf{V}}^{-1}\tilde{\mathbf{I}}_\mathcal{J}\mathbf{V}_L^{-1}
&= \setlength{\arraycolsep}{2.0pt}
\renewcommand{\arraystretch}{0.7}
\left[
\begin{matrix}
\alpha^{14} & \alpha^{13} & \alpha^{11} & \alpha^7 \\
\alpha^2 & \alpha^4 & \alpha^8 &\alpha \\
\alpha & \alpha^2 & \alpha^4 & \alpha^8 \\
1 & 1 & 1 & 1
\end{matrix}
\right]
\left[
\begin{matrix}
1 & \alpha^{-1} & \alpha^{-2} & \ldots & \alpha^{-13} & \alpha^{-14} \\
1 & \alpha^{-2} & \alpha^{-4} & \ldots & \alpha^{-11} & \alpha^{-13} \\
1 & \alpha^{-4} & \alpha^{-8} & \ldots & \alpha^{-7} & \alpha^{-11} \\
1 & \alpha^{-8} & \alpha^{-1} & \ldots & \alpha^{-14} & \alpha^{-7}
\end{matrix}
\right] \\
&= \setlength{\arraycolsep}{2.0pt}
\renewcommand{\arraystretch}{0.7}\left[
\begin{matrix}
1 & 1 & 1 & 1 & 0 & 1 & 0 & 1 & 1 & 0 & 0 & 1 & 0 & 0 & 0 \\
0 & 0 & 0 & 1 & 1 & 1 & 1 & 0 & 1 & 0 & 1 & 1 & 0 & 0 & 1 \\
0 & 0 & 1 & 1 & 1 & 1 & 0 & 1 & 0 & 1 & 1 & 0 & 0 & 1 & 0 \\
0 & 1 & 1 & 1 & 1 & 0 & 1 & 0 & 1 & 1 & 0 & 0 & 1 & 0 & 0 \\
\end{matrix}
\right].
\end{align*}
Justified by the above lemma, $\mathbf{G}_s$ is defined over GF($2$), so it is a candidate for the source encoding matrix. The next theorem further proves that $\mathbf{G}_s$ is indeed a desired one.

\begin{theorem}
\label{Thm:Connection_Circular-shift_Scalar}
Equipped with the source encoding matrix $\mathbf{G}_s = \mathbf{I}_\omega\otimes\mathbf{G}$, the circular-shift linear code $(\mathbf{K}_{d,e})$ constructed by (\ref{eqn:scalar-to-vector-circular-shift}) from a scalar linear code $(k_{d,e}(\alpha))$ qualifies as a $(J, L)$ linear solution.
\begin{proof}
This is continuation of the proof of Theorem \ref{thm:rank-relation}, with the additional $\mathbf{G}_s$ taken into account. We shall show that for every receiver $t$,
\begin{equation}
\label{eqn:proof_Connection_Circular-shift_Scalar}
\mathrm{rank}(\mathbf{G}_s[\mathbf{F}_e]_{e\in \mathrm{In}(t)}) = \sum\nolimits_{j \in \mathcal{J}}\mathrm{rank}([\mathbf{f}_e(\alpha^j)]_{e\in \mathrm{In}(t)}),
\end{equation}
which yields $\mathrm{rank}(\mathbf{G}_s[\mathbf{F}_e]_{e\in \mathrm{In}(t)}) = \omega J$, so that the code $(\mathbf{K}_{d,e})$ is a $(J, L)$ circular-shift linear solution by definiton. The proof of (\ref{eqn:proof_Connection_Circular-shift_Scalar}) is provided in Appendix-\ref{appendix:proof_thm_Connection_Circular-shift_Scalar}.
\end{proof}
\end{theorem}

\noindent \textbf{Remark}. The source encoding matrix $\mathbf{G}_s$ defined in (\ref{eqn:G_definition}) is not the unique one to turn the code $(\mathbf{K}_{d,e})$ into a $(J, L)$ linear solution, but it is a nontrivially and carefully designed one such that it applies to $(\mathbf{K}_{d,e})$ constructed from an arbitrary scalar linear code $(k_{d,e}(\alpha))$ by (\ref{eqn:scalar-to-vector-circular-shift}). One may wonder whether the simpler matrix $\mathbf{I}_\omega\otimes\tilde{\mathbf{I}}_\mathcal{J}$ can also be used as a source encoding matrix, for the reason that when $L$ is a prime with primitive root $2$ and $\mathcal{J} = \{1, 2, \ldots, L-1\}$, it becomes exactly the one adopted in \cite{Tang_Sun_Circular-shift_LNC} for the constructed $(L-1,L)$ circular-shift linear solution. %
We remark here that $\mathbf{I}_\omega\otimes\tilde{\mathbf{I}}_\mathcal{J}$ is insufficient to be a source encoding matrix for general odd $L$, as illustrated in the next example.

\begin{example}
Assume $\omega =2$, $L = 7$, and $[\mathbf{f}_e(x)]_{e\in \mathrm{In}(t)} = \left[\begin{matrix} 1 & 1 \\ 0 & 1+x+x^2+x^4 \end{matrix}\right]$ for some receiver $t$. In this case, when the primitive $7^{th}$ root of unity $\alpha \in \mathrm{GF}(2^3)$ is selected subject to $\alpha+\alpha^2+\alpha^4 = 1+\alpha^3+\alpha^5+\alpha^6 = 0$, we have
\[
\mathrm{rank}\left([\mathbf{f}_e(\alpha^j)]_{e\in \mathrm{In}(t)}\right) = \left\{
\begin{matrix}
1 & ~~\mathrm{when}~j \in \{0, 3, 5, 6\} \\
2 & \mathrm{when}~j \in \{1, 2, 4\}
\end{matrix}
\right.
\]
Correspondingly, set $\mathcal{J} = \{1, 2, 4\}$ and $\tilde{\mathbf{I}}_\mathcal{J} = \setlength{\arraycolsep}{2.0pt}
\renewcommand{\arraystretch}{0.7}
\left[\begin{matrix}
0 & 1 & 0 & 0 & 0 & 0 & 0 \\
0 & 0 & 1 & 0 & 0 & 0 & 0 \\
0 & 0 & 0 & 0 & 1 & 0 & 0
\end{matrix}\right]$. By (\ref{eqn:G_definition}),
\[
\mathbf{G} = \tilde{\mathbf{V}}^{-1}\tilde{\mathbf{I}}_\mathcal{J}\mathbf{V}_L^{-1}
= \setlength{\arraycolsep}{2.0pt}
\renewcommand{\arraystretch}{0.7}
\left[
\begin{matrix}
1 & \alpha & \alpha^2 \\
1 & \alpha^2 & \alpha^4 \\
1 & \alpha^4 & \alpha
\end{matrix}
\right]^{-1}
\left[
\begin{matrix}
1 & \alpha^{-1} & \alpha^{-2} & \ldots & \alpha^{-6} \\
1 & \alpha^{-2} & \alpha^{-4} & \ldots & \alpha^{-5} \\
1 & \alpha^{-4} & \alpha^{-1} & \ldots & \alpha^{-3}
\end{matrix}
\right]
= \left[\begin{matrix} 1 & 1 & 1 & 0 & 1 & 0 & 0 \\
0 & 0 & 1 & 1 & 1 & 0 & 1 \\
 0& 1 & 1 & 1 & 0 & 1 & 0 \end{matrix}\right],
\]
and it can be checked that
\[
\mathrm{rank}\left((\mathbf{I}_2\otimes\mathbf{G})
[\mathbf{f}_e(\mathbf{C}_7)]_{e\in \mathrm{In}(t)}\right) = 6.
\]
In contrast,
\[
\mathrm{rank}\left((\mathbf{I}_2\otimes\tilde{\mathbf{I}}_\mathcal{J})
[\mathbf{f}_e(\mathbf{C}_7)]_{e\in \mathrm{In}(t)}\right) = 5,
\]
so it is impossible for receiver $t$ to recover all $6$ source binary data symbols if $\mathbf{I}_2\otimes\tilde{\mathbf{I}}_\mathcal{J}$ is set as the source encoding matrix. \hfill $\blacksquare$
\end{example}

\subsection{Discussion on Code Rate $J/L$}
Consider an arbitrary scalar linear code $(k_{d,e}(\alpha))$ and the $L$-dimensional circular-shift linear code $(\mathbf{K}_{d,e})$ constructed by (\ref{eqn:scalar-to-vector-circular-shift}) from $(k_{d,e}(\alpha))$. According to Theorem \ref{Thm:Connection_Circular-shift_Scalar}, equipped with the source encoding matrix $\mathbf{G}_s$, $(\mathbf{K}_{d,e})$ is a circular-shift linear solution at rate $J/L$, where $J$ refers to the number of scalar linear codes $(k_{d,e}(\alpha^j))$, $0 \leq j \leq L-1$, that qualify to be a solution. As $(k_{d,e}(\alpha))$ itself may not be a linear solution, the code rate of the constructed $(\mathbf{K}_{d,e})$ needs to be calculated case by case. However, when determining the exact $J$, we need not check whether $(k_{d,e}(\alpha^j))$ is a solution for every $0 \leq j \leq L -1$. We now introduce an easier way to calculate $J/L$, by just checking whether $(k_{d,e}(\alpha^j))$ qualifies to be a linear solution with $j$ selected from a subset of $\{0, 1, \ldots, L-1\}$. For this goal, we need to recall the concept of cyclotomic polynomials, which will also be exploited in the subsequent sections.

Write
\begin{equation}
\label{eqn:set_R_coprime_to_L}
R = \{1 \leq r \leq L-1: \gcd(r, L) = 1\}.
\end{equation}
Denote by $\phi(L)$ the Euler's totient function of $L$, so $\phi(L) = |R|$. The $L^{th}$ cyclotomic polynomial over GF($2$) is
\[
Q_L(x) = \prod\nolimits_{r\in R}(x - \alpha^r).
\]
When $L$ is a prime with primitive root $2$, $Q_L(x)$ itself is an irreducible polynomial over GF($2$). For general odd $L$, the following lemma will be useful.

\begin{lemma}
\label{lemma:field_operation}
For a positive odd integer $L$, $\phi(L)$ is divisible by $m_L$. The cyclotomic polynomial $Q_L(x)$ factors into $\frac{\phi(L)}{m_L}$ irreducible polynomials $f_1(x), f_2(x), \ldots, f_{\phi(L)/m_L}(x)$ over GF($2$) of the same degree $m_L$.
\begin{proof}
See, for example, Theorem 2.47 in \cite{Lidl_Finite_Fields}.
\end{proof}
\end{lemma}

Because $\alpha$ is a primitive $L^{th}$ root of unity, the $L$ roots of $x^L - 1$ are exactly $\alpha^{j}$, $0 \leq j \leq L-1$. On the other hand, as $x^L - 1 = \prod_{L'|L} Q_{L'}(x)$ (See, e.g., Theorem 2.45 in \cite{Lidl_Finite_Fields}),
Lemma \ref{lemma:field_operation} implies that $x^L - 1$ factors into $\sum_{L'|L} \frac{\phi(L')}{m_{L'}}$ irreducible polynomials over GF($2$). For each irreducible polynomial that divides $x^L - 1$ and is of degree $m_{L'}$, if it has $\alpha^j$ as a root, then its $m_{L'}$ roots consist of $\alpha^{j2^l}$, $0 \leq l \leq m_{L'}-1$. Thus, the set $\{0, 1, 2, \ldots, L-1\}$ of integers can be partitioned into $\sum_{L'|L} \frac{\phi(L')}{m_{L'}}$ disjoint sets, each of which can be expressed as $\{j2^l: 0 \leq l \leq m_{L'}-1 \}$ modulo $L$ for some $0 \leq j \leq L-1$ and $L' | L$. %
As a consequence of Lemma \ref{lemma:scalar_linear_solution_induce_another_scalar}, given a scalar linear code $(k_{d,e}(\alpha))$, in order to check whether $(k_{d,e}(\alpha^j))$ qualifies to be a scalar linear solution for all $0 \leq j \leq L-1$, it suffices to only check the cases that $j$ is equal to exactly one (arbitrary) representative integer in each of the $\sum_{L'|L} \frac{\phi(L')}{m_{L'}}$ disjoint sets.

Though there is not an explicit characterization on the code rate $J/L$ of $(\mathbf{K}_{d,e})$ due to the generality of the considered scalar linear code $(k_{d,e}(\alpha))$, if $(k_{d,e}(\alpha))$ qualifies to be a linear solution, then the code rate $J/L$ of $(\mathbf{K}_{d,e})$ is at least $m_L$. This is because $\alpha, \alpha^2, \ldots, \alpha^{2^{m_L-1}}$ comprise the $m_L$ roots of an irreducible polynomial that divides $Q_L(x)$, so that at least $m_L$ scalar linear codes $(k_{d,e}(\alpha^{2^j}))$, $0 \leq j \leq m_L-1$ qualify to be a solution by Lemma \ref{lemma:scalar_linear_solution_induce_another_scalar}. In the next section, we shall further discuss how to construct a circular-shift linear solution at a higher rate $\phi(L)/L$, by properly designing $(k_{d,e}(\alpha))$ such that $(k_{d,e}(\alpha^{r}))$ is a scalar linear solution for all $r \in R$.

\section{Explicit Construction of a Circular-shift Linear Solution}
\subsection{Existence of a Circular-shift Linear Solution}
In the previous section, we have introduced a general method to map an arbitrary scalar linear code over GF($2^{m_L}$) to an $L$-dimensional circular-shift linear solution, but there is no explicit characterization on the code rate and the degree of $(\mathbf{K}_{d,e})$. %
In this section, we proceed to introduce the construction of an $L$-dimensional circular-shift linear solution at rate $\phi(L)/L$ of an arbitrary degree $\delta$.

For $1 \leq \delta \leq L-1$, denote by
\begin{equation}
\label{eqn:K_delta_x}
\mathcal{K}^{(x)}_\delta = \left\{\sum\nolimits_{j=0}^{L-1}a_jx^j: a_j \in \{0, 1\}, \sum\nolimits_{j = 0}^{L-1} a_j \leq \delta \right\}
\end{equation}
the set of polynomials over GF($2$) of degree at most $L-1$ and with at most $\delta$ nonzero terms. Analogously, for $0 \leq i \leq L-1$, write
\begin{align*}
\mathcal{K}^{(\alpha^i)}_\delta &= \left\{k(\alpha^i): k(x) \in \mathcal{K}^{(x)}_\delta\right\} = \left\{\sum\nolimits_{j=0}^{L-1}a_j\alpha^{ij}: a_j \in \{0, 1\}, \sum\nolimits_{j = 0}^{L-1} a_j \leq \delta \right\},
\end{align*}
that is, every element in $\mathcal{K}^{(\alpha^i)}_\delta$ corresponds to evaluation of a polynomial in $\mathcal{K}^{(x)}_\delta$ by setting $x = \alpha^i$. Note that for general odd $L$, it is possible to have two distinct polynomials $k_1(x), k_2(x) \in  \mathcal{K}^{(x)}_\delta$ subject to $k_1(\alpha^i) = k_2(\alpha^i)$. Hence, $\mathcal{K}^{(\alpha^i)}_\delta$ is a \emph{multiset} instead of a set. Moreover, when $\delta = m_L$, all elements in GF($2^{m_L}$) are contained in $\mathcal{K}^{(\alpha)}_\delta$, because $\{1, \alpha, \alpha^2, \ldots, \alpha^{m_L-1}\}$ forms a polynomial basis of GF($2^{m_L}$). Denote by $K_\delta$ the number of distinct elements in $\mathcal{K}^{(\alpha)}_\delta$.%

Based on Lemma \ref{lemma:field_operation}, we can obtain the following lemma, which plays a key role to prove the existence of a $(\phi(L), L)$ circular-shift linear solution of degree $\delta$ on a multicast network for block length $L$ subject to a constraint. Recall that as defined in $(\ref{eqn:set_R_coprime_to_L})$, $R$ consists of all integers between $1$ and $L-1$ that are coprime with $L$.

\begin{lemma}
\label{lemma:Schwartz_Zippel_Generalized}
Let $L$ be an odd integer, and $g(x_1, x_2, \ldots, x_n)$ a non-zero multivariate polynomial of degree at most $D$ in every $x_j$ over GF($2^{m_L}$). When $\frac{m_L}{\phi(L)}K_\delta  > D$, there exist $k_1(x), k_2(x), \ldots, k_n(x) \in \mathcal{K}_\delta^{(x)}$ such that the evaluation
\[
g(k_1(\alpha^r), k_2(\alpha^r), \ldots, k_n(\alpha^r)) \neq 0
\]
holds for all $r \in R$.
\begin{proof}
Please refer to Appendix-\ref{appendix:proof_lemma_Schwartz_Zippel_Generalized}.
\end{proof}
\end{lemma}

\begin{theorem}
\label{Thm:Linear_Solution_Existence_Odd_Block_Length}
Consider a multicast network with the set $T$ of receivers, an odd integer $L$, and degree $\delta$ with the associated set $\mathcal{K}_\delta^{(x)}$ defined in (\ref{eqn:K_delta_x}). When $\frac{m_L}{\phi(L)}K_\delta > |T|$, there exists a $(\phi(L), L)$ circular-shift linear solution of degree $\delta$.
\begin{proof}
We need to show that when $\frac{m_L}{\phi(L)}K_\delta > |T|$, there exists an assignment of $k_{d,e}(x) \in \mathcal{K}_\delta^{(x)}$ to every adjacent pair $(d,e)$, such that for all $r \in R$, the scalar linear code $(k_{d,e}(\alpha^r))$ over GF($2^{m_L}$) is a linear solution. This is because the circular-shift linear code $(\mathbf{K}_{d,e})$ constructed from such $(k_{d,e}(\alpha))$ by (\ref{eqn:scalar-to-vector-circular-shift}) is of degree $\delta$, and when it is equipped with the source encoding matrix $\mathbf{G}_s = \mathbf{I}_\omega\otimes(\tilde{\mathbf{V}}^{-1} \tilde{\mathbf{I}}_\mathcal{J}\mathbf{V}_L^{-1})$, where $\mathcal{J}$ is set to be $R$, it is a $(\phi(L), L)$ circular-shift linear solution according to Theorem \ref{Thm:Connection_Circular-shift_Scalar}.

Assign every adjacent pair $(d, e)$ an indeterminate $x_{d,e}$. Under the classical framework of LNC in \cite{KoetterMedard03} and the observation in \cite{HoKoetter03}, every multicast network can be associated with a polynomial, denoted by $F(\ast)$, over GF($2^{m_L}$) in indeterminates $\{x_{d,e}: \mathrm{adjacent~pair}~(d,e)\}$ such that
\begin{itemize}
\item the degree of $F(\ast)$ in every $x_{d,e}$ is at most $|T|$.
\item a scalar linear code $(k_{d,e})$ is a linear solution if and only if the evaluation of $F(\ast)$ by setting $x_{d,e} = k_{d,e}$ is a nonzero element in GF($2^{m_L}$).
\end{itemize}
When $\frac{m_L}{\phi(L)}K_\delta > |T|$, by Lemma \ref{lemma:Schwartz_Zippel_Generalized}, there exists an assignment of $k_{d,e}(x) \in \mathcal{K}_\delta^{(x)}$ for every adjacent pair $(d,e)$ such that the evaluation of $F(\ast)$ by setting $x_{d,e} = k_{d,e}(\alpha^r)$ is nonzero for all $r \in R$. Under such an assignment, $(k_{d,e}(\alpha^r))$ is a scalar linear solution for all $r \in R$.
\end{proof}
\end{theorem}

By proving the existence of such a scalar linear code $(k_{d,e}(\alpha))$ that $(k_{d,e}(\alpha^{r}))$ is a scalar linear solution for all $r \in R$, we have shown the existence of a $(\phi(L), L)$ circular-shift linear solution $(\mathbf{K}_{d,e})$. Write $L$ as the unique factorization $p_1^{l_1}p_2^{l_2}\ldots p_r^{l_r}$ where $p_1, p_2, \ldots, p_r$ are distinct primes and $l_1, l_2, \ldots, l_r$ are positive integers. Then, $\phi(L) = (p^{l_1}_1-p^{l_1-1}_1)(p^{l_2}_2-p^{l_2-1}_2)\ldots (p^{l_r}_r-p^{l_r-1}_r)$ and the code rate of $(\mathbf{K}_{d,e})$ can be expressed as
\begin{equation}
\label{eqn:rate_phi_L_expression}
\frac{\phi(L)}{L} = \left(1-\frac{1}{p_1}\right)\left(1-\frac{1}{p_2}\right)\ldots\left(1-\frac{1}{p_r}\right).
\end{equation}
Thus, under the consideration of Theorem \ref{Thm:Linear_Solution_Existence_Odd_Block_Length}, in order to obtain a circular-shift linear solution at a relatively higher rate, it would be better to select $L$ as a prime power.

\subsection{Efficient Construction}
Given a subset $F$ of a finite field GF($q$) with $|F| \geq |T|$, a well-known efficient algorithm was proposed in \cite{Jaggi05}, by a flow path approach, to construct a scalar linear solution over GF($q$) with all local encoding kernels belonging to $F$. In this subsection, we shall demonstrate that by the flow path approach, a $(\phi(L), L)$ circular-shift linear solution can also be efficiently constructed.

Adopt the same notation as in the previous subsection and assume $L$ is an odd integer. Justified by Theorem \ref{Thm:Connection_Circular-shift_Scalar}, in order to efficiently construct a $(\phi(L), L)$ circular-shift linear solution of degree $\delta$, it suffices to efficiently assign $k_{d,e}(x)\in \mathcal{K}_\delta^{(x)}$ for every adjacent pair $(d,e)$ such that for each $r \in R$, the scalar linear code $(k_{d,e}(\alpha^r))$ over GF($2^{m_L}$) is a linear solution. %

Denote by $C_j$, $1 \leq j \leq \frac{\phi(L)}{m_L}$, the cyclotomic cosets modulo $L$, and by $r_j$ an arbitrary entry in $C_j$. Thus, $R = \bigcup_{1 \leq j \leq \frac{\phi(L)}{m_L}} C_j$, and Lemma \ref{lemma:scalar_linear_solution_induce_another_scalar} asserts that $(k_{d,e}(\alpha^{r_j}))$ forming a scalar linear solution implies that $(k_{d,e}(\alpha^{r}))$ forms a scalar linear solution for all $r \in C_j$. Consequently, the task to efficiently assign $k_{d,e}(x)\in \mathcal{K}_\delta^{(x)}$ such that $(k_{d,e}(\alpha^r))$ is a scalar linear solution for each $r \in R$ can be further reduced to efficiently assign $k_{d,e}(x)\in \mathcal{K}_\delta^{(x)}$ such that $(k_{d,e}(\alpha^{r_j}))$ is a scalar linear solution for each $1 \leq j \leq \frac{\phi(L)}{m_L}$, which can be done by the next algorithm.

\begin{algorithm}
\label{alg:efficient_construction}
Assume $\lfloor\frac{m_L}{\phi(L)}K_\delta\rfloor > |T|$. As initialization, set $[\mathbf{f}_e(x)]_{e\in \mathrm{Out}(s)} = \mathbf{I}_\omega$, and for each receiver $t$,
\begin{itemize}
\item associate an arbitrary collection $\wp_t$ of $\omega$ edge-disjoint paths starting from $\mathrm{Out}(s)$ and ending at $\mathrm{In}(t)$;
\item set $I_{t} = \mathrm{Out}(s)$;
\item define $\omega\frac{\phi(L)}{m_L}$ $\omega$-dimensional vectors $\mathbf{w}_{t,e',j}$ (over GF($2^{m_L}$)) subject to $[\mathbf{w}_{t,e',j}]_{e' \in I_t} = \mathbf{I}_\omega$ for all $1 \leq j \leq \phi(L)/m_L$.
\end{itemize}
For every non-source node $v$, according to a topological order, perform procedures 1)-4) below for every $e \in \mathrm{Out}(v)$ to assign $k_{d_i,e}(x) \in \mathcal{K}_\delta^{(x)}$, compute $\mathbf{f}_e(x)$, and update $I_t$ and $\mathbf{w}_{t,e',j}$, so that the following two invariants always hold for all $t \in T$ and $1 \leq j \leq \phi(L)/m_L$:
\begin{align}
\label{eqn:alg_rank_invariant}
   \mathrm{rank}([\mathbf{f}_{e'}(\alpha^{r_j})]_{e'\in I_{t}}) &= \omega,\\
\label{eqn:alg_definition_vector_w}
   \mathbf{f}_{e'}(\alpha^{r_j})^\mathrm{T}\mathbf{w}_{t,e',j} = 1,
   \mathbf{f}_{d'}(\alpha^{r_j})^\mathrm{T}\mathbf{w}_{t,e',j} &= 0, \forall e' \in I_t, d' \in I_{t}\backslash\{e'\}
\end{align}
\begin{enumerate}
\item For each $d \in \mathrm{In}(v)$, denote by $T_d$ the set of such receivers $t$ that the adjacent pair $(d, e)$ is on some path in $\wp_{t}$. If $|T_d| = 0$, then set $k_{d,e}(x) = 0$.
\item Let $\{d_1, \ldots, d_l\}$ denote the subset of $\mathrm{In}(v)$ with $|T_{d_i}| > 0$ for all $1 \leq i \leq l$. Note that a receiver $t$ can only appear in at most one set $T_{d_i}$, $1 \leq i \leq l$.\footnote{This is because all paths in $\wp_t$ can contain at most one among adjacent pairs $(d, e), d \in \mathrm{In}(t)$.}
\item If $l = 0$, then end the current iteration for $e$. Otherwise, for $i = 1$, set $k_{d_i, e}(x) = 1$ and define $\mathbf{f}(x) = \mathbf{f}_{d_i}(x)$. For $2 \leq i \leq l$, iteratively assign $k_{d_i,e}(x) \in \mathcal{K}_\delta^{(x)}$ and update $\mathbf{f}(x)$ in the following way so as to keep
    \begin{equation}
    \label{eqn:algorithm_fx_invariance}
    \mathbf{f}(\alpha^{r_j})^\mathrm{T}\mathbf{w}_{t,d_{i'},j} \neq 0 ~ \forall 1 \leq i' \leq i, t \in T_{d_{i'}}, 1 \leq j \leq \phi(L)/m_L.
    \end{equation}%
    after every iteration $i$.
    \begin{itemize}
    \item
     If $\mathbf{f}(\alpha^{r_j})^\mathrm{T}\mathbf{w}_{t,d_i,j} \neq 0$ for all $t \in T_{d_i}$ and $1 \leq j \leq \frac{\phi(L)}{m_L}$, then set $k_{d_i, e}(x) = 0$, keep $\mathbf{f}(x)$ unchanged, and end the current iteration on $i$.
    \item
      Otherwise, for $1 \leq j \leq \frac{\phi(L)}{m_L}$, define $\mathcal{A}_j$ as
      \noindent
      \begin{align*}
      \mathcal{A}_j = &\Big\{-\frac{\mathbf{f}_{d_i}(\alpha^{r_j})^\mathrm{T}\mathbf{w}_{t,d_{i'},j}} {\mathbf{f}(\alpha^{r_j})^\mathrm{T}\mathbf{w}_{t,d_{i'},j}}: t \in T_{d_{i'}},1\leq i' < i\Big\} \bigcup \\
      &\Big\{-\frac{\mathbf{f}_{d_i}(\alpha^{r_j})^\mathrm{T}\mathbf{w}_{t,d_{i},j}} {\mathbf{f}(\alpha^{r_j})^\mathrm{T}\mathbf{w}_{t,d_{i},j}}:  t \in T_{d_i}, \mathbf{f}(\alpha^{r_j})^\mathrm{T}\mathbf{w}_{t,d_i,j} \neq 0 \Big\}
      \end{align*}%
      Note that under the inductive assumption (\ref{eqn:algorithm_fx_invariance}) up to iteration $i-1$, whose correctness will be justified in Proposition \ref{Prop:Algorithm_justification}, such $\mathcal{A}_j$ is well defined.
    \item Set $k_{d_i,e}(x)$ to be a polynomial in $\mathcal{K}_\delta^{(x)}$ such that for all $1 \leq j \leq \frac{\phi(L)}{m_L}$,
        \begin{equation}
        \label{eqn:algorim_LEK_selection}
        k_{d_i,e}(\alpha^{r_j}) \neq 0,~ k_{d_i,e}(\alpha^{r_j})^{-1} \notin \mathcal{A}_j.
        \end{equation}
        As to be justified in Proposition \ref{Prop:Algorithm_justification}, such $k_{d_i,e}(x)$ can always be selected.
    \item Reset $\mathbf{f}(x)$ to be $\mathbf{f}(x)+k_{d_i,e}(x)\mathbf{f}_{d_i}(x)$.
    \end{itemize}
    \item Set $\mathbf{f}_e(x)$ as $\mathbf{f}(x)$. For every $t \in T_{d_i}$, $1 \leq i \leq l$, replace $I_t$ by $I_t\cup\{e\}\backslash\{d_i\}$, further define
        \begin{equation}
        \label{eqn:alg_new_w_e_define}
        \mathbf{w}_{t,e,j} = (\mathbf{f}_e(\alpha^{r_j})^{\mathrm{T}}\mathbf{w}_{t,d_i,j})^{-1}\mathbf{w}_{t,d_i,j}, 1 \leq j \leq \phi(L)/m_L,
        \end{equation}
        and update $\mathbf{w}_{t,d',j}$, where $d' \in I_t \backslash \{e\}$, $1 \leq j \leq \phi(L)/m_L$, as
        \begin{equation}
        \label{eqn:alg_w_d_update}
        \mathbf{w}_{t,d',j} - (\mathbf{f}_e(\alpha^{r_j})^{\mathrm{T}}\mathbf{w}_{t,d',j})\mathbf{w}_{t,e,j}.       \end{equation}
        The iteration for edge $e$ completes, and as justified by Proposition \ref{Prop:Algorithm_justification},  (\ref{eqn:alg_rank_invariant}) and (\ref{eqn:alg_definition_vector_w}) keeps correct.
  \end{enumerate}
After completion of the above procedures, $I_{t} = \mathrm{In}(t)$ for all $t \in T$, and $k_{d,e}(x) \in \mathcal{K}_\delta^{(x)}$ has been set for every adjacent pair $(d,e)$.  \hfill $\blacksquare$
\end{algorithm}

\begin{figure}[!t]
\centering
\scalebox{0.63}
{\includegraphics{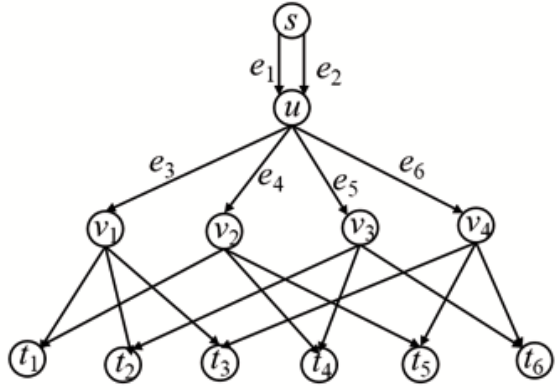}}
\caption{The classical (4,2)-Combination Network.}
\label{Combination_Network_Algorithm8}
\end{figure}

We shall next illustrate Algorithm \ref{alg:efficient_construction} based on the $(4, 2)$-Combination Network. The classical $(n, 2)$-Combination Network, $n \geq 4$, is a special multicast network consisting of four layers of nodes. The unique source node $s$ comprises the first layer, and its two outgoing edges lead to the layer-2 node $u$. There are $4$ nodes at the third layer, each of which is connected from $u$ by an edge. For every pair of layer-3 nodes, there is a bottom-layer receiver connected from them. There are total $\left( \begin{smallmatrix} n \\ 2 \end{smallmatrix} \right)$ receivers, each of which needs to recover the $2$ source data units generated by $s$. %

\begin{example}
Consider the $(4,2)$-Combination Network as depicted in Fig. \ref{Combination_Network_Algorithm8}.  Assume $L = 7$, so that $m_L = 3$ and $\phi(L)=6$. Now we shall adopt Algorithm \ref{alg:efficient_construction} to assign $k_{d,e}(x) \in \mathcal{K}_{1}^{(x)}$ for every adjacent pair $(d,e)$. Set $r_1 = 1$, $r_2 = 3$ so that they are in different cyclotomic cosets modulo $7$.

As initialization, prescribe
\begin{equation*}
\begin{matrix}
\wp_{t_1}=(e_1e_3e_{11},e_2e_4e_{21}), \wp_{t_2}=(e_1e_3e_{12},e_2e_5e_{32}) \\ \wp_{t_3}=(e_1e_3e_{13},e_2e_6e_{43}),
\wp_{t_4}=(e_1e_4e_{24},e_2e_5e_{34}) \\ \wp_{t_5}=(e_1e_4e_{25},e_2e_6e_{45}), \wp_{t_6}=(e_1e_5e_{36},e_2e_6e_{46}),
\end{matrix}
\end{equation*}
where $e_{ij}$ refers to the edge from node $v_i$ to receiver $t_j$. For every receiver $t \in T$, set $I_{t} = \mathrm{Out}(s)=\{e_1,e_2\}$ and $\mathbf{f}_{e_1}(x) = \mathbf{w}_{t, e_1, j} = [1~0]^{\mathrm{T}}, \mathbf{f}_{e_2}(x) = \mathbf{w}_{t, e_2, j} = [0~1]^{\mathrm{T}}, j \in \{1, 2\}$.

For node $u$, the algorithm will sequentially deal with its outgoing edges $e_3, \ldots, e_6$ as follows:
\begin{itemize}
\item Iteration for $e_3$. Step 1) yields $T_{e_1}=\{t_1,t_2,t_3\}$, $T_{e_2}=\phi$, and set $k_{e_2, e_3}(x) = 0$. Step 2) then defines $d_1 = e_1$. Step 3) sets $k_{d_1, e_3}(x) = 1$, $\mathbf{f}(x) = \mathbf{f}_{d_1}(x)$. After Step 4), the current iteration ends with the update $\mathbf{f}_{e_3}(x) = \mathbf{f}(x) = [1 ~ 0]^{\mathrm{T}}$, $I_{t_1} = I_{t_2} = I_{t_3} = \{e_3, e_2\}$, and
    \[
    \mathbf{w}_{t_1, e_3, j} = \mathbf{w}_{t_2, e_3, j} = \mathbf{w}_{t_3, e_3, j} = [1~0]^{\mathrm{T}},
    \mathbf{w}_{t_1, e_2, j} = \mathbf{w}_{t_2, e_2, j} = \mathbf{w}_{t_3, e_2, j} = [0~1]^{\mathrm{T}},
    \]
\item Iteration for $e_4$. Step 1) yields $T_{e_1}=\{t_4,t_5\}$, $T_{e_2}=\{t_1\}$. Step 2) defines $d_1 = e_1$, $d_2 = e_2$. In Step 3), for $i = 1$, the algorithm sets $k_{d_1, e_4}(x) = 1$ and $\mathbf{f}(x) = \mathbf{f}_{d_1}(x) = [1~0]^{\mathrm{T}}$. For $i = 2$, as $t_1 \in T_{d_2}$ and $\mathbf{f}(\alpha^{r_j})^\mathrm{T}\mathbf{w}_{t_1,d_2,j} = 0$, the algorithm needs proceed to obtain $\mathcal{A}_j = \{0 \}$, $j \in \{1, 2\}$. In order to satisfy (\ref{eqn:algorim_LEK_selection}), the algorithm can assign $k_{d_2, e_4}(x) = x$, and then reset $\mathbf{f}(x)$ to be $\mathbf{f}(x)+k_{d_2, e_4}(x)\mathbf{f}_{d_2}(x) = [1~x]^{\mathrm{T}}$. After Step 4), the current iteration ends with the update $\mathbf{f}_{e_4}(x) = \mathbf{f}(x) = [1 ~ x]^{\mathrm{T}}$, $I_{t_4} = I_{t_5} = \{e_4, e_2\}, I_{t_1} = \{e_3, e_4\}$, and
    \begin{align*}
    \mathbf{w}_{t_4, e_4, j} &= (\mathbf{f}_{e_4}(\alpha^{r_j})^{\mathrm{T}}\mathbf{w}_{t_4, d_1, j})^{-1} \mathbf{w}_{t_4, d_1, j} = [1~0]^{\mathrm{T}}, \\
    \mathbf{w}_{t_4, e_2, j} &= \mathbf{w}_{t_4, e_2, j} - (\mathbf{f}_{e_4}(\alpha^{r_j})^{\mathrm{T}}\mathbf{w}_{t_4, e_2, j})\mathbf{w}_{t_4, e_4, j} = [\alpha^{r_j} ~ 1]^{\mathrm{T}},
    \\
    \mathbf{w}_{t_5, e_4, j} &= \mathbf{w}_{t_4, e_4, j},
    \mathbf{w}_{t_5, e_2, j} = \mathbf{w}_{t_4, e_2, j}, \\
    \mathbf{w}_{t_1, e_4, j} &= (\mathbf{f}_{e_4}(\alpha^{r_j})^{\mathrm{T}}\mathbf{w}_{t_1, d_2, j})^{-1} \mathbf{w}_{t_1, d_2, j} = [0~\alpha^{-r_j}]^{\mathrm{T}},
    \\
    \mathbf{w}_{t_1, e_3, j} &= \mathbf{w}_{t_1, e_3, j} - (\mathbf{f}_{e_4}(\alpha^{r_j})^{\mathrm{T}}\mathbf{w}_{t_1, e_3, j})\mathbf{w}_{t_1, e_4, j} = [1~\alpha^{-r_j}]^{\mathrm{T}}.
    \end{align*}
\item Iteration for $e_5$. Step 1) yields $T_{e_1}=\{t_6\}$, $T_{e_2}=\{t_2, t_4\}$. Step 2) defines $d_1 = e_1$, $d_2 = e_2$. In Step 3), for $i = 1$, the algorithm sets $k_{d_1, e_5}(x) = 1$ and $\mathbf{f}(x) = \mathbf{f}_{d_1}(x) = [1~0]^{\mathrm{T}}$. For $i = 2$, as $t_2 \in T_{d_2}$ and $\mathbf{f}(\alpha^{r_j})^\mathrm{T}\mathbf{w}_{t_2,d_2,j} = 0$, the algorithm needs proceed to obtain
    \[
    \mathcal{A}_j = \Big\{-\frac{\mathbf{f}_{d_2}(\alpha^{r_j})^{\mathrm{T}}\mathbf{w}_{t_6,d_1,j}}
    {\mathbf{f}(\alpha^{r_j})^{\mathrm{T}}\mathbf{w}_{t_6,d_1,j}},
    -\frac{\mathbf{f}_{d_2}(\alpha^{r_j})^{\mathrm{T}}\mathbf{w}_{t_4,d_2,j}}
    {\mathbf{f}(\alpha^{r_j})^{\mathrm{T}}\mathbf{w}_{t_4,d_2,j}}
    \Big\}
    = \{0, \alpha^{-r_j} \}
    \]
    In order to satisfy (\ref{eqn:algorim_LEK_selection}), the algorithm can assign $k_{d_2, e_5}(x) = x^2$, and then reset $\mathbf{f}(x)$ to be $\mathbf{f}(x)+k_{d_2, e_5}(x)\mathbf{f}_{d_2}(x) = [1~x^2]^{\mathrm{T}}$. After Step 4), the current iteration ends with the update
    $\mathbf{f}_{e_5}(x) = \mathbf{f}(x) = [1 ~ x^2]^{\mathrm{T}}$, $I_{t_6} = \{e_5, e_2\}$, $I_{t_2} = \{e_3, e_5\}$, $I_{t_4} = \{e_4, e_5\}$, and
    \begin{align*}
    \mathbf{w}_{t_6, e_5, j} &= [1~0]^{\mathrm{T}}, \mathbf{w}_{t_6, e_2, j} = [\alpha^{2r_j}~1]^{\mathrm{T}}, \\
    \mathbf{w}_{t_2, e_5, j} &= [0~\alpha^{-2r_j}]^{\mathrm{T}},
    \mathbf{w}_{t_2, e_3, j} = [1~\alpha^{-2r_j}]^{\mathrm{T}}, \\
    \mathbf{w}_{t_4, e_5, j} &= \frac{1}{\alpha^{2r_j}+\alpha^{r_j}}[\alpha^{r_j}~1]^{\mathrm{T}},
    \mathbf{w}_{t_4, e_4, j} =     \frac{1}{\alpha^{2r_j}+\alpha^{r_j}}[\alpha^{2r_j}~1]^{\mathrm{T}},
    \end{align*}
\item Iteration for $e_6$. Step 1) yields $T_{e_2}=\{t_3,t_5,t_6\}$, $T_{e_1}=\phi$, and set $k_{e_1, e_6}(x) = 0$. Step 2) then defines $d_1 = e_2$. Step 3) sets $k_{d_1, e_6}(x) = 1$, $\mathbf{f}(x) = \mathbf{f}_{d_1}(x)$. After Step 4), the current iteration ends with the update $\mathbf{f}_{e_6}(x) = \mathbf{f}(x) = [0 ~ 1]^{\mathrm{T}}$, $I_{t_3} = \{e_3, e_6\}, I_{t_5} = \{e_4, e_6\}, I_{t_6} = \{e_5, e_6\}$, and
    \begin{align*}
    \mathbf{w}_{t_3, e_6, j} &= [0~1]^{\mathrm{T}}, \mathbf{w}_{t_3, e_3, j} = [1~0]^{\mathrm{T}}, \\
    \mathbf{w}_{t_5, e_6, j} &= [\alpha^{r_j}~1]^{\mathrm{T}},
    \mathbf{w}_{t_5, e_4, j} = [1~0]^{\mathrm{T}}, \\
    \mathbf{w}_{t_6, e_6, j} &= [\alpha^{2r_j}~1]^{\mathrm{T}}, \mathbf{w}_{t_6, e_5, j} = [1~0]^{\mathrm{T}}. \\
    \end{align*}
\end{itemize}
One may check that after the iteration on each of the edges $e_3, \ldots, e_6$ completes, (\ref{eqn:alg_rank_invariant}) and (\ref{eqn:alg_definition_vector_w}) always hold. For each node $v_i$, $1 \leq i \leq 4$, as its indegree is $1$ and every adjacent pair $(e_i, e_{ij})$ is on some path in a certain $\wp_t$, the algorithm will set $k_{e_i,e_{ij}}(x) = 1$. Up to now, every adjacent pair has been assigned a polynomial in $\mathcal{K}_1^{(x)}$. It can be readily checked that $(k_{d,e}(\alpha^{j}))$ qualifies as a scalar linear solution for all $1 \leq j \leq 6$. Subsequently, based on (\ref{eqn:scalar-to-vector-circular-shift}) and Theorem \ref{Thm:Connection_Circular-shift_Scalar}, a $(6, 7)$ circular shift-linear solution of degree $1$ can be constructed. \hfill $\blacksquare$
\end{example}

\begin{proposition}
\label{Prop:Algorithm_justification}
In Algorithm \ref{alg:efficient_construction}, after every iteration of Step 3), $k_{d_i,e}(x)$ can always be selected from $\mathcal{K}_\delta^{(x)}$ subject to (\ref{eqn:algorim_LEK_selection}), and condition (\ref{eqn:algorithm_fx_invariance}) always holds. %
In addition, when the iteration for an arbitrary edge completes, (\ref{eqn:alg_rank_invariant}) and (\ref{eqn:alg_definition_vector_w}) always hold. %
Consequently, based on the constructed $k_{d,e}(x) \in \mathcal{K}_\delta^{(x)}$ by Algorithm \ref{alg:efficient_construction}, $(k_{d,e}(\alpha^{r}))$ forms a scalar linear solution over GF($2^{m_L}$) for all $r \in R$.
\begin{proof}
Please refer to Appendix-\ref{appendix:proof_prop_algorithm_justification}.
\end{proof}
\end{proposition}

One may notice that in Theorem \ref{Thm:Linear_Solution_Existence_Odd_Block_Length}, $\frac{m_L}{\phi(L)}K_\delta > |T|$ is sufficient to guarantee the existence of a $(\phi(L), L)$ circular-shift linear solution of degree $\delta$, but the efficient construction of such a code by Algorithm \ref{alg:efficient_construction} requires $\lfloor \frac{m_L}{\phi(L)}K_\delta \rfloor > |T|$. This slight difference is a cost of selecting $k_{d_i,e}(x)$ subject to (\ref{eqn:algorim_LEK_selection}) in Step 3) of the algorithm, where every $k_{d_i,e}(x)$, once assigned, needs \emph{not} be updated any more. Such an easier manipulation on $k_{d_i,e}$ is new, and is different from the original flow path approach in \cite{Jaggi05}.

We next theoretically analyze the computational complexity of Algorithm \ref{alg:efficient_construction}. In the initialization step, for each receiver $t \in T$, it takes $\mathcal{O}(|E|\omega)$ operations to establish $\wp_t$ by the augmenting path approach. %
After initialization, Algorithm \ref{alg:efficient_construction} traverses every edge exactly once. In every iteration to deal with an edge, Step 3) requires $l \leq |T|$ iterations and in each iteration: %
i) it takes $\mathcal{O}(|T|\omega)$ operations to compute the values in every $\mathcal{A}_j$, $1 \leq j \leq \frac{\phi(L)}{m_L}$; %
ii) it takes at most $\mathcal{O}\left(\frac{\phi(L)^2}{m_L^2}|T|\right)$ operations to select $k_{d_i,e}(x)$ from $\mathcal{K}_\delta^{(x)}$ prescribed by $(\ref{eqn:algorim_LEK_selection})$, where computing the evaluation of $k_{d_i,e}(x)$ at $x = \alpha^{r_j}$ can be avoided by setting a mapping table from $\mathcal{K}_\delta^{(x)}$ to $\mathcal{K}_\delta^{(\alpha^j)}$ in advance. Step 4) requires to update at most $|T|\omega\frac{\phi(L)}{m_L}$ vectors $\mathbf{w}_{t,e',j}$, each of which involves $\mathcal{O}(\omega)$ operations. In summary, the computational complexity of the algorithm is $\mathcal{O}\left(|E||T|\omega + \frac{\phi(L)}{m_L}|E||T|^2\omega + \frac{\phi(L)^2}{m_L^2}|E||T|^2+|E||T|\omega^2\frac{\phi(L)}{m_L}\right)
= \mathcal{O}\left(\frac{\phi(L)}{m_L}|E||T|(\omega^2 + \omega|T| + \frac{\phi(L)}{m_L}|T|)\right)$.

In the practical design of a $(\phi(L), L)$ circular-shift linear solution of degree $\delta$ by Algorithm \ref{alg:efficient_construction}, the parameter $\phi(L)$ requires to be calculated in advance. As discussed at the end of the previous subsection, $\phi(L) = (p^{l_1}_1-p^{l_1-1}_1)(p^{l_2}_2-p^{l_2-1}_2)\ldots (p^{l_r}_r-p^{l_r-1}_r)$, where $p_1, \ldots, p_r$ are distinct prime divisors of $L$ and $p_1^{l_1}p_2^{l_2}\ldots p_r^{l_r}$ is the unique factorization of $L$. Thus, the complexity to calculate $\phi(L)$ for an arbitrary $L$ is essentially same as the unique factorization of $L$. Though the unique factorization of $L$ is known to have extremely high computational complexity for very large $L$, according to \cite{Number_Theory}, its computing cost is acceptable when $L$ is as moderately large as $10^{10}$. In addition, we can consider some particular $L$ such as power primes so that $\phi(L)$ can be easily computed.

We end this section by listing some design instances of a $(\phi(L), L)$ circular-shift linear solution of degree $\delta$.
\begin{itemize}
\item Assume $L$ is prime with primitive root $2$ and $1 \leq \delta \leq \frac{L-1}{2}$, so that $m_L = \phi(L) = L-1$ and all elements in $\mathcal{K}_\delta^{(\alpha)}$ are distinct. Thus, $K_\delta = \sum_{j=0}^\delta \left(\begin{smallmatrix} L-1 \\ j \end{smallmatrix}\right)$. When $K_\delta > |T|$, an $(L-1, L)$ circular-shift linear solution of degree $\delta$ can be efficiently constructed. This is the case considered in \cite{Tang_Sun_Circular-shift_LNC}.
\item Assume $\delta = 1$, so that $K_\delta = L+1$. When $\frac{L+1}{L-1}m_L > |T|$, a $(\phi(L), L)$ circular-shift linear solution of degree $\delta$ can be efficiently constructed.
\item Assume $\delta = m_L$, so that all elements in GF($2^{m_L}$) are contained in $\mathcal{K}_\delta^{(\alpha)}$ and $K_\delta = 2^{m_L}$. When $\frac{m_L}{\phi(L)}2^{m_L} > |T|$, a $(\phi(L), L)$ circular-shift linear solution of degree $\delta$ can be efficiently constructed.
\item Assume $L = p^l$, where $p$ is an odd prime, so that $\phi(L) = p^l-p^{l-1}$ and $m_L = m_pp^{l-1}$. When $\frac{m_p}{p-1}K_\delta > |T|$, a $(p^l-p^{l-1}, p^l)$ circular-shift linear solution of degree $\delta$ can be efficiently constructed.
\end{itemize}

Note that since $K_1$ is always larger than $\phi(L)$, when $m_L \geq |T|$, a $(\phi(L), L)$ circular-shift linear solution of degree $1$ can be efficiently constructed.

\section{Asymptotical Linear Solvability of Circular-shift LNC}
Circular-shift LNC has been proven insufficient to achieve the exact multicast capacity of some multicast networks \cite{Tang_Sun_Circular-shift_LNC}. Whether every multicast network is \emph{asymptotically circular-shift linearly solvable}, that is, for any $\epsilon > 0$, it has an $(L', L)$ circular-shift linear solution with $L'/L > 1 - \epsilon$, becomes a fundamental problem for theoretical study of circular-shift LNC. For the case that $L$ is a prime with primitive root $2$, the efficient construction of an $(L-1, L)$ circular-shift linear solution has been discussed \cite{Tang_Sun_Circular-shift_LNC}. However, whether there are infinitely many primes with primitive root $2$ is still unknown (See, e.g, \cite{Sloane_Prime_List}), so whether every multicast network is asymptotically circular-shift linearly solvable remains open. %
As an application of Theorem \ref{Thm:Linear_Solution_Existence_Odd_Block_Length}, which applies to a general odd block length $L$, every multicast network is known to have an $(L-1,L)$ circular-shift linear solution for an arbitrary prime $L$ with $m_L > |T|$, so consequently we are able to give an affirmative answer to this open problem.

\begin{theorem}
\label{Thm:Asymptotic_Linear_Solvability}
Every multicast network is asymptotically circular-shift linearly solvable.
\begin{proof}
Consider an arbitrary multicast network with the set $T$ of receivers. %
Let $\epsilon$ be an arbitrary positive value, and write $M = \max\{\lceil\frac{1}{\epsilon}\rceil, |T|\}$. For an arbitrary positive integer $m$, denote by $P_m$ the set of primes modulo which the multiplicative order of $2$ is equal to $m$. As $p$ divides $2^m - 1$ for each $p \in P_m$, $P_m$ contains finitely many primes, and thus so does $\bigcup_{m < M}P_m$. As there are infinitely many primes, there must exist a prime $L$ so that its multiplicative order $m_L \geq M$. For such $L$ with $\phi(L) = L-1$, according to Theorem \ref{Thm:Linear_Solution_Existence_Odd_Block_Length}, there exists an $(L-1, L)$ circular-shift linear solution. In addition, as $L > m_L > \frac{1}{\epsilon}$, $\frac{L-1}{L} > 1 - \epsilon$.
\end{proof}
\end{theorem}

\section{Summary and Concluding Remarks}
In the present paper, we formulated circular-shift linear network coding (LNC) for an arbitrary odd block length $L$, in the context of multicast networks. In particular, we introduced a method to induce an $L$-dimensional circular-shift linear solution (over GF($2$)) at rate $J/L$ from an arbitrary scalar linear code $(k_{d,e}(\alpha))$ over GF($2^{m_L}$), where $m_L$ is the multiplicative order of $2$ modulo $L$, $\alpha \in \mathrm{GF}(2^{m_L})$ is a primitive $L^{th}$ root of unity, and $J$ is the number of integers $j$ from $0$ to $L-1$ such that $(k_{d,e}(\alpha^{j}))$ is a scalar linear solution. %
Stemming from this connection, we further demonstrated, by both an existence proof argument and an efficient construction argument, that every multicast network has an $L$-dimensional circular-shift linear solution at rate $\phi(L)/L$, where $\phi(L)$ is the Euler's totient function of $L$. Furthermore, we proved an open conjecture proposed in \cite{Tang_Sun_Circular-shift_LNC} that every multicast network is asymptotically circular-shift linearly solvable. Potentially, the intrinsic connection between circular-shift LNC and scalar LNC established herein can be extended to general networks, and the present consideration of circular-shift LNC over GF($2$) can also be theoretically generalized to over GF($q$). We leave them as future work.

\appendix
\subsection{Proof of Theorem \ref{thm:rank-relation}}
\label{appendix:proof-rank-relation}
According to the classical framework in \cite{KoetterMedard03}, the global encoding kernels $\mathbf{f}_e(\alpha)$ incoming to $t$ can be expressed as
\[
[\mathbf{f}_e(\alpha)]_{e\in\mathrm{In}(t)} = \mathbf{A}(\alpha) \left(\mathbf{I}_{|E|-\omega}+\mathbf{K}(\alpha)+\mathbf{K}(\alpha)^2 +\ldots\right)\mathbf{B}(1).
\]
Here $\mathbf{A}(\alpha)$ and $\mathbf{K}(\alpha)$ respectively stand for the $\omega \times (|E|-\omega)$ matrix $[k_{d,e}(\alpha)]_{d \in \mathrm{Out}(s), e\notin \mathrm{Out}(s)}$ and the $(|E|-\omega) \times (|E|-\omega)$ matrix $[k_{d,e}(\alpha)]_{d, e\notin \mathrm{Out}(s)}$ for brevity,%
\footnote{$[k_{d,e}]_{d\in A, e\in B}$ is defined to be such an $|A|\times|B|$ matrix with rows and columns respectively indexed by $A \subseteq E$ and $B \subseteq E$ that the entry indexed by $(d,e)$ is equal to $k_{d,e}(\alpha)$.} %
and $\mathbf{B}(1)$ is an $(|E| - \omega)\times|\mathrm{In}(t)|$ index matrix of which the unique nonzero entry $1$ in every column corresponds to an edge in $\mathrm{In}(t)$. Via replacing $\alpha$ in $\mathbf{A}(\alpha)$, $\mathbf{K}(\alpha)$ by the cyclic permutation matrix $\mathbf{C}_L$, and replacing $1$ in $\mathbf{B}(1)$ by $\mathbf{I}_L$, we have
\[
[\mathbf{F}_e]_{e\in\mathrm{In}(t)} = \mathbf{A}(\mathbf{C}_L) (\mathbf{I}_{(|E|-\omega)L}+\mathbf{K}(\mathbf{C}_L)+\mathbf{K}(\mathbf{C}_L)^2 +\ldots)\mathbf{B}(\mathbf{I}_L).
\]%
Based on (\ref{eqn:cyclic_matrix_decomposition}), %
which applies to an arbitrary positive odd integer $L$,
\begin{align*}
&\mathbf{A}(\mathbf{C}_L) = (\mathbf{I}_\omega\otimes\mathbf{V}_L)\cdot\mathbf{A}(\mathbf{\Lambda}_\alpha)\cdot(\mathbf{I}_{|E|-\omega}\otimes\mathbf{V}_L^{-1}) , \\
&\mathbf{K}(\mathbf{C}_L)^j = \mathbf{K}(\mathbf{V}_L\mathbf{\Lambda}_\alpha\mathbf{V}_L^{-1})^j =(\mathbf{I}_{|E|-\omega}\otimes\mathbf{V}_L)\cdot\mathbf{K}(\mathbf{\Lambda}_\alpha)^j\cdot(\mathbf{I}_{|E|-\omega}\otimes\mathbf{V}_L^{-1}),\\
&\mathbf{B}_t(\mathbf{I}_L)= (\mathbf{I}_{|E|-\omega}\otimes\mathbf{V}_L) \cdot \mathbf{B}_t(\mathbf{I}_L)\cdot (\mathbf{I}_{\omega}\otimes\mathbf{V}_L^{-1}),
\end{align*}
where $\otimes$ represents the Kronecker product. Thus,
\begin{equation}
\label{eqn:Fe_decomposition}
[\mathbf{F}_e]_{e\in\mathrm{In}(t)} = (\mathbf{I}_\omega\otimes\mathbf{V}_L)\cdot\mathbf{M}\cdot(\mathbf{I}_{\omega}\otimes\mathbf{V}_L^{-1}),
\end{equation}
where
\[
\mathbf{M} = \mathbf{A}(\mathbf{\Lambda}_\alpha)
\left(\mathbf{I}_{(|E|-\omega)L}+\mathbf{K}(\mathbf{\Lambda}_\alpha)+\mathbf{K}(\mathbf{\Lambda}_\alpha)^2 + \ldots \right)\mathbf{B}(\mathbf{I}_L).
\]

As $(\mathbf{I}_\omega\otimes\mathbf{V}_L)$ and $(\mathbf{I}_\omega\otimes\mathbf{V}_L)^{-1}$ are full rank $\omega L$,
\[
\mathrm{rank}([\mathbf{F}_e]_{e\in\mathrm{In}(t)}) = \mathrm{rank}(\mathbf{M}).
\]

Since all of $\mathbf{A}(\mathbf{\Lambda}_\alpha)$, $\mathbf{K}(\mathbf{\Lambda}_\alpha)^j$, and $\mathbf{B}(\mathbf{I}_L)$ can be regarded as a block matrix with every block entry to be an $L \times L$ diagonal matrix, so is $\mathbf{M}$. %
Thus, we can rearrange the rows and columns in $\mathbf{M}$ to form a new matrix $\tilde{\mathbf{M}}$ as follows. Let $\mathbf{P}$ denote the $\omega L \times \omega L$ permutation matrix that can be written in the block form %
$\left[\begin{smallmatrix} \mathbf{J}_{1,1} & \ldots & \mathbf{J}_{1, \omega} \\ \vdots & \ddots & \vdots \\ \mathbf{J}_{L,1} & \ldots & \mathbf{J}_{L, \omega} \end{smallmatrix}\right]$, %
where every block $\mathbf{J}_{i,j}$, $1 \leq i \leq L$, $1 \leq j \leq \omega$, is an $\omega\times L$ matrix with the only nonzero entry $1$ located at row $j$ and column $i$. Set $\tilde{\mathbf{M}} = \mathbf{P}\mathbf{M}\mathbf{P}^T$. It can be checked that
\begin{equation}
\label{eqn:M_tilde_expression}
\tilde{\mathbf{M}} = \setlength{\arraycolsep}{3.0pt}
\renewcommand{\arraystretch}{0.7}
\left[\begin{matrix}
\mathbf{M}_0 & \mathbf{0} & \mathbf{0} \\
\mathbf{0} & \ddots & \mathbf{0} \\
\mathbf{0} & \mathbf{0} & \mathbf{M}_{L-1}
\end{matrix}\right],
\end{equation}
where
\begin{equation}
\label{eqn:M_j_as_fe_j_expression}
\mathbf{M}_j = \mathbf{A}(\alpha^j) \left(\mathbf{I}_{|E|-\omega}+\mathbf{K}(\alpha^j)+\mathbf{K}(\alpha^j)^2 +\ldots\right)\mathbf{B}(1).
\end{equation}
Under the expression in (\ref{eqn:M_j_as_fe_j_expression}), it turns out that
\begin{equation}
\label{eqn:M_j_equal_f_e_alpha_j}
\mathbf{M}_j = [\mathbf{f}_e(\alpha^j)]_{e \in \mathrm{In}(t)}.
\end{equation}
So we have
\begin{align*}
\mathrm{rank}([\mathbf{F}_e]_{e\in\mathrm{In}(t)}) &= \mathrm{rank}(\mathbf{M}) = \mathrm{rank}(\tilde{\mathbf{M}}) = \sum\nolimits_{j=0}^{L-1}\mathrm{rank}([\mathbf{f}_e(\alpha^j)]_{e \in \mathrm{In}(t)}).
\end{align*}


\subsection{Proof of Lemma \ref{lemma:G_over_GF2}}
\label{appendix:G_over_GF2}
Denote by $\tilde{\mathbf{I}}_J$ the $J\times L$ matrix obtained from $\mathbf{I}_L$ by restricting to the first $J$ rows. Thus,
\begin{equation}
\label{eqn:V_tilde_expression_in_Lemma}
\tilde{\mathbf{V}} = \tilde{\mathbf{I}}_\mathcal{J}\mathbf{V}_L\tilde{\mathbf{I}}_J^T.
\end{equation}
Based on (\ref{eqn:Vandermonde_p}) and (\ref{eqn:Vandermond_L_inverse}), it can be easily seen that $\mathbf{V}_L^{-1}$ is a column permutation of $\mathbf{V}_L$. Thus, in order to show that every entry in $\mathbf{G} = \tilde{\mathbf{V}}^{-1} \tilde{\mathbf{I}}_\mathcal{J}\mathbf{V}^{-1}_L$ is in GF($2$), it is equivalent to show that every entry in $\tilde{\mathbf{V}}^{-1}\tilde{\mathbf{I}}_\mathcal{J}\mathbf{V}_L$ is in GF($2$).

As a complement of $\tilde{\mathbf{I}}_\mathcal{J}$, denote by $\tilde{\mathbf{I}}_{\mathcal{J}^c}$ the $(L-J)\times L$ matrix obtained from $\mathbf{I}_L$ by deleting the $(j+1)^{st}$ row whenever $j \in \mathcal{J}$. Since $\tilde{\mathbf{I}}_\mathcal{J}^T\tilde{\mathbf{I}}_\mathcal{J}
+\tilde{\mathbf{I}}_{\mathcal{J}^c}^T\tilde{\mathbf{I}}_{\mathcal{J}^c} = \mathbf{I}_L$,
\[
\mathbf{V}_L^{-1}(\tilde{\mathbf{I}}_{\mathcal{J}}^T
\tilde{\mathbf{I}}_{\mathcal{J}} + \tilde{\mathbf{I}}_{\mathcal{J}^c}^T
\tilde{\mathbf{I}}_{\mathcal{J}^c})\mathbf{V}_L = \mathbf{I}_L.
\]
In addition, since $\tilde{\mathbf{I}}_J\tilde{\mathbf{I}}_J^T = \mathbf{I}_J$,
\begin{equation}
\label{eqn:GF2_lemma_proof_matrix_manipulation}
\tilde{\mathbf{I}}_J \mathbf{V}_L^{-1} \tilde{\mathbf{I}}_\mathcal{J}^T
\tilde{\mathbf{I}}_\mathcal{J} \mathbf{V}_L \tilde{\mathbf{I}}_J^T +
\tilde{\mathbf{I}}_J \mathbf{V}_L^{-1} \tilde{\mathbf{I}}_{\mathcal{J}^c}^T
\tilde{\mathbf{I}}_{\mathcal{J}^c} \mathbf{V}_L \tilde{\mathbf{I}}_J^T
= \tilde{\mathbf{I}}_J
(\mathbf{V}_L^{-1}(\tilde{\mathbf{I}}_{\mathcal{J}}^T
\tilde{\mathbf{I}}_{\mathcal{J}} + \tilde{\mathbf{I}}_{\mathcal{J}^c}^T
\tilde{\mathbf{I}}_{\mathcal{J}^c})\mathbf{V}_L)
\tilde{\mathbf{I}}_J^T
= \mathbf{I}_J.
\end{equation}

For simplicity, write
\[
\mathbf{U}_1 = \tilde{\mathbf{I}}_J \mathbf{V}_L^{-1} \tilde{\mathbf{I}}_\mathcal{J}^T,~
\mathbf{U}_2 = \tilde{\mathbf{I}}_J \mathbf{V}_L^{-1} \tilde{\mathbf{I}}_{\mathcal{J}^c}^T \tilde{\mathbf{I}}_{\mathcal{J}^c} \mathbf{V}_L \tilde{\mathbf{I}}_J^T.
\]
Together with (\ref{eqn:V_tilde_expression_in_Lemma}), Eq. (\ref{eqn:GF2_lemma_proof_matrix_manipulation}) can be written as
$\mathbf{U}_1\tilde{\mathbf{V}} + \mathbf{U}_2 = \mathbf{I}_J$. %
Because $\mathbf{U}_1^T$ and $\tilde{\mathbf{V}}$ can be respectively regarded as a Vandermonde matrix generated by $\alpha^{-j}, j \in \mathcal{J}$ and by $\alpha^j, j\in \mathcal{J}$, they are invertible, and so is $\mathbf{I}_J + \mathbf{U}_2$. Thus,

\[
\tilde{\mathbf{V}}^{-1} = (\mathbf{I}_J + \mathbf{U}_2)^{-1}\mathbf{U}_1,
\]
and
\[
\tilde{\mathbf{V}}^{-1}\tilde{\mathbf{I}}_\mathcal{J}\mathbf{V}_L = (\mathbf{I}_J + \mathbf{U}_2)^{-1} (\tilde{\mathbf{I}}_J \mathbf{V}_L^{-1} \tilde{\mathbf{I}}_\mathcal{J}^T) \tilde{\mathbf{I}}_\mathcal{J}\mathbf{V}_L.
\]
As the inverse of a matrix over GF($2$) is also over GF($2$), it turns out that in order to show that $\tilde{\mathbf{V}}^{-1}\tilde{\mathbf{I}}_\mathcal{J}\mathbf{V}_L$ is a matrix over GF($2$), it suffices to show that both $\mathbf{V}_L^{-1} \tilde{\mathbf{I}}_\mathcal{J}^T \tilde{\mathbf{I}}_\mathcal{J} \mathbf{V}_L$ %
and
$\mathbf{U}_2$ are over GF($2$).

Among integers $0, 1, \ldots, L-1$, label the ones in $\mathcal{J}$ as $j_1, \ldots, j_{J}$, and the ones not in $\mathcal{J}$ as $j_{J+1}, \ldots, j_{L}$, both in an ascending order. We have
\[
\mathbf{V}_L^{-1} \tilde{\mathbf{I}}_\mathcal{J}^T
= \left[
\begin{matrix}
1 \\
\alpha^{-j_l} \\
\alpha^{-2j_l} \\
\vdots \\
\alpha^{-(L-1)j_l}
\end{matrix}
\right]_{1 \leq l \leq J},~
\tilde{\mathbf{I}}_\mathcal{J}\mathbf{V}_L
= \left[
\begin{matrix}
1 \\
\alpha^{j_l} \\
\alpha^{2j_l} \\
\vdots \\
\alpha^{(L-1)j_l}
\end{matrix}
\right]_{1 \leq l \leq J}^T.
\]
Thus, for all $1 \leq l_1, l_2 \leq L$, the $(l_1, l_2)^{th}$ entry in $\mathbf{V}_L^{-1} \tilde{\mathbf{I}}_\mathcal{J}^T \tilde{\mathbf{I}}_\mathcal{J} \mathbf{V}_L$ can be written as
\[
\sum\nolimits_{l=1}^J \alpha^{-l_1 j_l}\alpha^{l_2 j_l} = \sum\nolimits_{l=1}^J \alpha^{(l_2-l_1)j_l}.
\]
Because Lemma \ref{lemma:scalar_linear_solution_induce_another_scalar} implies that $\mathcal{J} = \{j_1, \ldots, j_J\}$ is closed under multiplication by $2$ modulo $L$, we have
\[
\left(\sum\nolimits_{l=1}^J \alpha^{(l_2-l_1)j_l}\right)^2
= \sum\nolimits_{l=1}^J \alpha^{2(l_2-l_1)j_l}
= \sum\nolimits_{l=1}^J \alpha^{(l_2-l_1)j_l},
\]
and so $\sum_{l=1}^J \alpha^{(l_2-l_1)j_l} \in \mathrm{GF}(2)$, \emph{i.e.}, every entry in $\mathbf{V}_L^{-1} \tilde{\mathbf{I}}_\mathcal{J}^T \tilde{\mathbf{I}}_\mathcal{J} \mathbf{V}_L$ belongs to GF($2$). %

Similarly, for all $1 \leq l_1, l_2 \leq L$, the $(l_1, l_2)^{th}$ entry in $\mathbf{V}_L^{-1} \tilde{\mathbf{I}}_{\mathcal{J}^c}^T \tilde{\mathbf{I}}_{\mathcal{J}^c} \mathbf{V}_L$ can be expressed as $\sum_{l=J+1}^L \alpha^{(l_2-l_1)j_l}$. Since both $\{0, 1, \ldots, L-1\}$ and $\mathcal{J}$ are closed under multiplication by $2$ modulo $L$, $\mathcal{J}^c = \{j_{J+1}, j_{J+2}, \ldots, j_L\} = \{0, 1, \ldots, L-1\}\backslash\mathcal{J}$ is also closed under multiplication by $2$ modulo $L$. Thus, $\left(\sum_{l=J+1}^L \alpha^{(l_2-l_1)j_l}\right)^2 = \sum_{l=J+1}^L \alpha^{(l_2-l_1)j_l}$ and so $\sum_{l=J+1}^L \alpha^{(l_2-l_1)j_l} \in \mathrm{GF}(2)$, \emph{i.e.}, every entry in $\mathbf{V}_L^{-1} \tilde{\mathbf{I}}_{\mathcal{J}^c}^T \tilde{\mathbf{I}}_{\mathcal{J}^c} \mathbf{V}_L$ belongs to GF($2$). As $\mathbf{U}_2 = \tilde{\mathbf{I}}_J(\mathbf{V}_L^{-1} \tilde{\mathbf{I}}_{\mathcal{J}^c}^T \tilde{\mathbf{I}}_{\mathcal{J}^c} \mathbf{V}_L)\tilde{\mathbf{I}}_J^T$, $\mathbf{U}_2$ is over GF($2$).

\subsection{Proof of Theorem \ref{Thm:Connection_Circular-shift_Scalar}}
\label{appendix:proof_thm_Connection_Circular-shift_Scalar}
It remains to prove (\ref{eqn:proof_Connection_Circular-shift_Scalar}). Follow the same argument as in the proof of Theorem \ref{thm:rank-relation} (refer to Appendix-\ref{appendix:proof-rank-relation}) till Eq. (\ref{eqn:M_j_equal_f_e_alpha_j}). For a receiver $t$, by (\ref{eqn:Fe_decomposition}), (\ref{eqn:M_tilde_expression}) and (\ref{eqn:M_j_equal_f_e_alpha_j}), we have
\[
[\mathbf{F}_e]_{e\in\mathrm{In}(t)} = (\mathbf{I}_\omega\otimes\mathbf{V}_L)\mathbf{P}^T
\tilde{\mathbf{M}}
\mathbf{P}(\mathbf{I}_{\omega}\otimes\mathbf{V}_L^{-1}),
\]
where $\tilde{\mathbf{M}} = \left[\begin{smallmatrix}
\mathbf{M}_0 & \mathbf{0} & \mathbf{0} \\
\mathbf{0} & \ddots & \mathbf{0} \\
\mathbf{0} & \mathbf{0} & \mathbf{M}_{L-1}
\end{smallmatrix}\right]$, and $\mathbf{M}_j = [\mathbf{f}_e(\alpha^j)]_{e\in \mathrm{In}(t)}$.

By (\ref{eqn:G_definition}), $\mathbf{G}_s = \mathbf{I}_\omega \otimes (\tilde{\mathbf{V}}^{-1}\tilde{\mathbf{I}}_\mathcal{J}\mathbf{V}_L^{-1})$. Similar to the definition of the permutation matrix $\mathbf{P}$, define $\mathbf{Q}$ as the $\omega J \times \omega J$ permutation matrix that can be written in the block form $\left[\begin{smallmatrix} \mathbf{J}_{1,1} & \ldots & \mathbf{J}_{1, \omega} \\ \vdots & \ddots & \vdots \\ \mathbf{J}_{J,1} & \ldots & \mathbf{J}_{J, \omega} \end{smallmatrix}\right]$, where every block $\mathbf{J}_{i,j}$, $1 \leq i \leq J$, $1 \leq j \leq \omega$, is an $\omega\times J$ matrix with the only nonzero entry $1$ located at row $j$ and column $i$. As $\mathbf{Q}^T\mathbf{Q} = \mathbf{I}_{\omega J}$,
\[
\mathbf{G}_s = \mathbf{I}_\omega\otimes(\tilde{\mathbf{V}}^{-1}\tilde{\mathbf{I}}_\mathcal{J}\mathbf{V}_L^{-1})
= (\mathbf{I}_\omega \otimes \tilde{\mathbf{V}}^{-1})\mathbf{Q}^T \mathbf{Q}
(\mathbf{I}_\omega\otimes(\tilde{\mathbf{I}}_\mathcal{J}\mathbf{V}_L^{-1})).
\]
and thus
\[
\mathbf{G}_s[\mathbf{F}_e]_{e\in\mathrm{In}(t)}
= (\mathbf{I}_\omega \otimes \tilde{\mathbf{V}}^{-1})\mathbf{Q}^T \mathbf{Q}
(\mathbf{I}_\omega\otimes\tilde{\mathbf{I}}_\mathcal{J})\mathbf{P}^T
\tilde{\mathbf{M}}
\mathbf{P}(\mathbf{I}_{\omega}\otimes\mathbf{V}_L^{-1}).
\]
Since the square matrix $(\mathbf{I}_\omega \otimes \tilde{\mathbf{V}}^{-1})\mathbf{Q}^T$ is full rank $\omega J$ and the square matrix $\mathbf{P}(\mathbf{I}_{\omega}\otimes\mathbf{V}_L^{-1})$ is full rank $\omega L$,
\begin{align*}
\mathrm{rank}(\mathbf{G}_s[\mathbf{F}_e]_{e\in\mathrm{In}(t)})
&=\mathrm{rank}\left(\mathbf{Q}(\mathbf{I}_\omega\otimes\tilde{\mathbf{I}}_\mathcal{J})\mathbf{P}^T
\tilde{\mathbf{M}}\right) \\
&=\mathrm{rank}\left((\tilde{\mathbf{I}}_\mathcal{J}\otimes\mathbf{I}_\omega)\tilde{\mathbf{M}}\right)
= \mathrm{rank}\left(\sum\nolimits_{j \in \mathcal{J}} [\mathbf{f}_e(\alpha^j)]_{e\in\mathrm{In}(t)}\right).
\end{align*}

\subsection{Proof of Lemma \ref{lemma:Schwartz_Zippel_Generalized}
\label{appendix:proof_lemma_Schwartz_Zippel_Generalized}}
According to Lemma \ref{lemma:field_operation}, the cyclotomic polynomial $Q_L(x) = f_1(x)\cdots f_{\phi(L)/m_L}(x)$, where $f_j(x)$, $1 \leq j \leq d$, is an irreducible polynomial over GF($2$) of degree $m_L$. Thus, for every $f_j(x)$, the exponents of the $m_L$ roots, expressed as powers of $\alpha$, constitute a cyclotomic coset $\{r, 2r, \ldots, 2^{m_L-1}r\}$ modulo $L$ for some $r \in R$, and $R$ can be partitioned into $\frac{\phi(L)}{m_L}$ distinct cyclotomic cosets modulo $L$.

Denote by $C_j$, $1 \leq j \leq \frac{\phi(L)}{m_L}$, the cyclotomic cosets modulo $L$ such that $f_j(x) = \prod_{r \in C_j}(x-\alpha^r)$, and by $r_j$ an arbitrary entry in $C_j$. If there exist $k_1(x), k_2(x), \ldots, k_n(x) \in \mathcal{K}_\delta^{(x)}$ subject to $g(k_1(\alpha^{r_j}), \ldots, k_n(\alpha^{r_j})) \neq 0$ for some $1 \leq j \leq d$, then for every $r' \in C_j$, which can be written as $2^lr_j$ modulo $L$ for some $1 \leq l \leq m_L$,
\begin{align*}
g(k_1(\alpha^{r'}), \ldots, k_n(\alpha^{r'})) & = g(k_1(\alpha^{2^lr_j}), \ldots, k_n(\alpha^{2^lr_j}))
=g(k_1(\alpha^{r_j})^{2^l}, \ldots, k_n(\alpha^{r_j})^{2^l}) \\
&=g(k_1(\alpha^{r_j}), \ldots, k_n(\alpha^{r_j}))^{2^l} \neq 0.
\end{align*}
In addition, as $R = \bigcup_{1\leq j \leq \frac{\phi(L)}{m_L}}C_j$, in order to show the lemma, it suffices to show the existence of $k_1(x), k_2(x), \ldots, k_n(x) \in \mathcal{K}_\delta^{(x)}$ such that
\begin{equation}
\label{eqn:Generalized-S-Z-Lemma-nonzero-evaluation}
\prod\nolimits_{1\leq j \leq \frac{\phi(L)}{m_L}} g(k_1(\alpha^{r_j}), k_2(\alpha^{r_j}), \ldots, k_n(\alpha^{r_j})) \neq 0.
\end{equation}

Note that for each $r \in R$, as $r$ is coprime with $L$, we have $\{\alpha, \alpha^2, \ldots, \alpha^L\} = \{\alpha^{r}, \alpha^{2r}, \ldots, \alpha^{Lr}\}$. Hence, the mapping $\psi: \mathcal{K}_\delta^{(\alpha)} \rightarrow \mathcal{K}_\delta^{(\alpha^r)}$ defined by $\psi(k(\alpha)) = k(\alpha^r)$ is a bijection, and $\mathcal{K}_\delta^{(\alpha)} = \mathcal{K}_\delta^{(\alpha^r)}$.
Consequently, there exist $k_1(\alpha^r), k_2(\alpha^r), \ldots, k_n(\alpha^r) \in \mathcal{K}_\delta^{(\alpha^r)}$ with $g(k_1(\alpha^{r}), k_2(\alpha^{r}), \ldots, k_n(\alpha^{r})) = 0$ if and only if there exists $k_1'(\alpha), k_2'(\alpha), \ldots, k_n'(\alpha) \in \mathcal{K}_\delta^{(\alpha)}$ with $g(k_1'(\alpha), k_2'(\alpha), \ldots, k_n'(\alpha)) = 0$.

As $g(x_1, \ldots, x_n)$ has degree at most $D$ in every $x_j$, by the Schwartz-Zippel lemma (See, e.g., \cite{Schwartz-Zippel}), it has at most $DK_\delta^{n-1}$ roots over $\mathcal{K}_\delta'^{(\alpha)}$, where $\mathcal{K}_\delta'^{(\alpha^i)}$ refers to the set containing all different elements in $\mathcal{K}_\delta^{(\alpha^i)}$ for , $0 \leq i \leq L-1$. %
Since $\mathcal{K}_\delta'^{(\alpha^{r_j})} = \mathcal{K}_\delta'^{(\alpha)}$ and $|\mathcal{K}_\delta'^{(\alpha^{r_j})}| = K_\delta$ for all $1 \leq j \leq \frac{\phi(L)}{m_L}$, by taking a union bound, we conclude that there are at most $\frac{\phi(L)}{m_L}DK_\delta^{n-1}$ possible choices of $k_1(\alpha), k_2(\alpha), \ldots, k_n(\alpha) \in \mathcal{K}_\delta'^{(\alpha)}$, where $k_i(\alpha)$ is the evaluation of some $k_i(x) \in \mathcal{K}_{\delta}^{(x)}$ by setting $x = \alpha$, such that
\[
\prod\nolimits_{1\leq j \leq \frac{\phi(L)}{m_L}} g(k_1(\alpha^{r_j}), k_2(\alpha^{r_j}), \ldots, k_n(\alpha^{r_j})) = 0.
\]%
Consequently, when $K_\delta > \frac{\phi(L)}{m_L}D$, \emph{i.e.}, $\frac{m_L K_\delta }{\phi(L)} > D$, there must exist  $k_1(x), k_2(x), \ldots, k_n(x) \in \mathcal{K}_\delta^{(x)}$ satisfying $g(k_{1}(\alpha^{r_j}), k_{2}(\alpha^{r_j}), \ldots, k_{j}(\alpha^{r_j})) \neq 0$ for all $1 \leq j \leq \frac{\phi(L)}{m_L}$, \emph{i.e.}, Eq. (\ref{eqn:Generalized-S-Z-Lemma-nonzero-evaluation}) obeys.

\subsection{Proof of Proposition \ref{Prop:Algorithm_justification}}
\label{appendix:proof_prop_algorithm_justification}
As initialization, (\ref{eqn:alg_rank_invariant}) and (\ref{eqn:alg_definition_vector_w}) obviously hold. Consider the case that the algorithm starts to deal with edge $e\in \mathrm{Out}(v)$ for some non-source node $v$, and assume that (\ref{eqn:alg_rank_invariant}) and (\ref{eqn:alg_definition_vector_w}) are correct with respect to the current setting of $I_t$, $t\in T$.

In Step 3), for every iteration $1 \leq i \leq l$, we first show that $k_{d_i,e}(x)$ can always be set subject to (\ref{eqn:algorim_LEK_selection}). By definition, $k_{d,e}(\alpha^{r_j}) \in \mathcal{K}_{\delta}^{(\alpha^{r_j})}$. Denote by $\mathcal{K}_\delta'^{(\alpha^{r_j})}$ the set containing all distinct elements in $\mathcal{K}_\delta^{(\alpha^{r_j})}$. As a receiver can only appear in at most one $T_{d_{i'}}$, $1 \leq i' \leq i$,
\[
|\mathcal{A}_j| \leq |\cup_{1 \leq i' \leq i}T_{d_{i'}}| \leq |T|.
\]
Hence, there are at most $|T|$ nonzero values in $\mathcal{K}_\delta'^{(\alpha^{r_j})}$ whose multiplicative inverses belong to $\mathcal{A}_j$. %
Since the algorithm will directly set $k_{d_i,e}(x) = 0$ for the case $\mathbf{f}(\alpha^{r_j})^\mathrm{T}\mathbf{w}_{t,d_{i'},j} \neq 0$ for all $t \in \bigcup\nolimits_{1 \leq i' \leq i} T_{d_{i'}}$ and $1 \leq j \leq \frac{\phi(L)}{m_L}$, as long as $\mathcal{A}_j$ needs to be involved for selecting $k_{d_i,e}(x)$,
\[
\left|\bigcup\nolimits_{1\leq j \leq \frac{\phi(L)}{m_L}}\mathcal{A}_j\right|
\leq \frac{\phi(L)}{m_L}|\mathcal{A}_j| \leq \frac{\phi(L)}{m_L}|T|-1.
\] %

It has been argued in the proof of Lemma \ref{lemma:Schwartz_Zippel_Generalized} that $\mathcal{K}_\delta'^{(\alpha)} = \mathcal{K}_\delta'^{(\alpha^{r_j})}$. %
Under the union bound, there are at most $\frac{\phi(L)}{m_L}(|T|+1)-1$ elements in the set
\begin{align*}
\big\{k(\alpha) \in \mathcal{K}_\delta'^{(\alpha)}: k(x) \in \mathcal{K}_\delta^{(x)}, ~\mathrm{either}~& k(\alpha^{r_j})  =0~\mathrm{or}~k(\alpha^{r_j})^{-1} \in \mathcal{A}_j ~\mathrm{for~some~} 1\leq j \leq \phi(L)/m_L \big\}.
\end{align*}
As it is assumed at the beginning of Algorithm \ref{alg:efficient_construction} that $\lfloor\frac{m_L}{\phi(L)}K_\delta\rfloor > |T|$, \emph{i.e.}, $K_\delta (= |\mathcal{K}_\delta'^{(\alpha)}|) > \frac{\phi(L)}{m_L}(|T|+1)-1$, there must exist $k_{d_i,e}(x) \in \mathcal{K}_\delta^{(x)}$ such that (\ref{eqn:algorim_LEK_selection}) holds for all $1 \leq j \leq \frac{\phi(L)}{m_L}$.

We next prove the correctness of (\ref{eqn:algorithm_fx_invariance}). After iteration $i = 1$, $\mathbf{f}(x)$ is set equal to $\mathbf{f}_{d_1}(x)$, so the inductive assumption (\ref{eqn:alg_definition_vector_w}) implies  $\mathbf{f}(\alpha^{r_j})^{\mathrm{T}}\mathbf{w}_{t,d_1,j} = 1 \neq 0$ for all $t \in T_{d_1}$ and $1 \leq j \leq \phi(L)/m_L$, \emph{i.e.}, condition (\ref{eqn:algorithm_fx_invariance}) holds. Inductively, assume (\ref{eqn:algorithm_fx_invariance}) is correct up to iteration $i-1$, where $1 < i \leq l$ and $\mathbf{f}(x)$ is obtained after iteration $i-1$. %
After iteration $i$, consider an arbitrary $1 \leq j \leq \frac{\phi(L)}{m_L}$. For the case $t \in T_{d_i}$, $\mathbf{f}(\alpha^{r_j})^\mathrm{T}\mathbf{w}_{t,d_i,j} \neq 0$,
\begin{equation}
\label{eqn:justify_f_w_d_neq_0_1}
\left(\mathbf{f}(\alpha^{r_j}) + k_{d_i,e}(\alpha^{r_j})\mathbf{f}_{d_i}(\alpha^{r_j})\right)^\mathrm{T}\mathbf{w}_{t,d_i,j}
= \mathbf{f}(\alpha^{r_j})^\mathrm{T}\mathbf{w}_{t,d_i,j} + k_{d_i,e}(\alpha^{r_j})\mathbf{f}_{d_i}(\alpha^{r_j})^\mathrm{T}\mathbf{w}_{t,d_i,j} \neq 0,
\end{equation}
and for the case $t \in \bigcup_{1\leq i' < i}T_{d_{i'}}$,
\begin{equation}
\label{eqn:justify_f_w_d_neq_0_2}
\left(\mathbf{f}(\alpha^{r_j}) + k_{d_i,e}(\alpha^{r_j})\mathbf{f}_{d_i}(\alpha^{r_j})\right)^\mathrm{T}\mathbf{w}_{t,d_{i'},j}
= \mathbf{f}(\alpha^{r_j})^\mathrm{T}\mathbf{w}_{t,d_{i'},j} + k_{d_i,e}(\alpha^{r_j})\mathbf{f}_{d_i}(\alpha^{r_j})^\mathrm{T}\mathbf{w}_{t,d_{i'},j} \neq 0,
\end{equation}
where the last inequality in both (\ref{eqn:justify_f_w_d_neq_0_1}) and (\ref{eqn:justify_f_w_d_neq_0_2}) is due to the constraints $k_{d_i,e}(\alpha^{r_j}) \neq 0$ and $k_{d_i,e}(\alpha^{r_j})^{-1} \notin \mathcal{A}_j$ in (\ref{eqn:algorim_LEK_selection}). %
In addition, for the case $t \in T_{d_i}$, $\mathbf{f}(\alpha^{r_j})^\mathrm{T}\mathbf{w}_{t,d_i,j} = 0$,
\[
\left(\mathbf{f}(\alpha^{r_j}) + k_{d_i,e}(\alpha^{r_j})\mathbf{f}_{d_i}(\alpha^{r_j})\right)^\mathrm{T}\mathbf{w}_{t,d_i,j} = k_{d_i,e}(\alpha^{r_j})\mathbf{f}_{d_i}(\alpha^{r_j})^\mathrm{T}\mathbf{w}_{t,d_i,j}
= k_{d_i,e}(\alpha^{r_j}) \neq 0,
\]
where the second equality is due to $\mathbf{f}_{d_i}(\alpha^{r_j})^\mathrm{T}\mathbf{w}_{t,d_i,j} = 1$ by the inductive assumption (\ref{eqn:alg_definition_vector_w}). We have thus verified that when $\mathbf{f}(x)$ is replaced by $\mathbf{f}(x)+k_{d_i,e}(x)\mathbf{f}_{d_i}(x)$ after iteration $i$, (\ref{eqn:algorithm_fx_invariance}) is still correct.

In Step 4), after $\mathbf{f}_e(x)$ is set to $\mathbf{f}(x)$, consider an arbitrary receiver $t \in T_{d_i}$, $1 \leq i \leq l$, and assume that $I_t$ has been replaced by $I_t\cup\{e\}\backslash\{d_i\}$. Since $\mathbf{f}_e(\alpha^{r_j})^{\mathrm{T}}\mathbf{w}_{t,d_i,j} = \mathbf{f}(\alpha^{r_j})^{\mathrm{T}}\mathbf{w}_{t,d_i,j} \neq 0$ by (\ref{eqn:algorithm_fx_invariance}), $\mathbf{w}_{t,e,j}$ in (\ref{eqn:alg_new_w_e_define}) is well defined and obviously
\begin{equation}
\label{eqn:f_e_w_e_equal_1}
\mathbf{f}_e(\alpha^{r_j})^{\mathrm{T}}\mathbf{w}_{t,e,j} = 1.
\end{equation}
In addition, for $d' \in I_t\backslash\{e\}$,
\begin{equation}
\label{eqn:f_e_w_d'_equal_0}
\mathbf{f}_{d'}(\alpha^{r_j})^{\mathrm{T}}\mathbf{w}_{t,e,j} = (\mathbf{f}_e(\alpha^{r_j})^{\mathrm{T}}\mathbf{w}_{t,d_i,j})^{-1}\mathbf{f}_{d'}(\alpha^{r_j})^{\mathrm{T}}\mathbf{w}_{t,d_i,j} = 0,
\end{equation}
where the last inequality is due to the inductive assumption (\ref{eqn:alg_definition_vector_w}). This further implies that
\begin{align*}
&\mathbf{f}_{d'}(\alpha^{r_j})^{\mathrm{T}}(\mathbf{w}_{t,d',j} - (\mathbf{f}_e(\alpha^{r_j})^{\mathrm{T}}\mathbf{w}_{t,d',j})\mathbf{w}_{t,e,j}) \\
= &\mathbf{f}_{d'}(\alpha^{r_j})^{\mathrm{T}}\mathbf{w}_{t,d',j} - (\mathbf{f}_e(\alpha^{r_j})^{\mathrm{T}}\mathbf{w}_{t,d',j})\mathbf{f}_{d'}(\alpha^{r_j})\mathbf{w}_{t,e,j}
= 1-0 = 1.
\end{align*}
Last,
\[
\mathbf{f}_{e}(\alpha^{r_j})^{\mathrm{T}}(\mathbf{w}_{t,d',j} - (\mathbf{f}_{e}(\alpha^{r_j})^{\mathrm{T}}\mathbf{w}_{t,d',j})\mathbf{w}_{t,e,j}) =(\mathbf{f}_{e}(\alpha^{r_j})^{\mathrm{T}}\mathbf{w}_{t,d',j})(1 - \mathbf{f}_{e}(\alpha^{r_j})^{\mathrm{T}}\mathbf{w}_{t,e,j}) = 0.
\]
We have thus verified that after the computation of $\mathbf{w}_{t,e,j}$ in (\ref{eqn:alg_new_w_e_define}) and the update of $\mathbf{w}_{t,d',j}$ as $\mathbf{w}_{t,d',j} - (\mathbf{f}_{e}(\alpha^{r_j})^{\mathrm{T}}\mathbf{w}_{t,d',j})\mathbf{w}_{t,e,j}$ in (\ref{eqn:alg_w_d_update}), Eq. (\ref{eqn:alg_definition_vector_w}) keeps correct. By the inductive assumption (\ref{eqn:alg_rank_invariant}), $\mathbf{f}_{d'}(\alpha^{r_j}), d' \in I_t\backslash\{e\}$ are linearly independent. Because of (\ref{eqn:f_e_w_e_equal_1}) and (\ref{eqn:f_e_w_d'_equal_0}), Lemma 5 in \cite{Jaggi05} can then be applied here to assert that $\mathbf{f}(\alpha^{r_j})$ is linearly independent of $\mathbf{f}_{d'}(\alpha^{r_j}), d'\in I_{t}\backslash\{e\}$. We have proved the correctness of (\ref{eqn:alg_rank_invariant}) after the iteration for edge $e$ completes.

When the algorithm terminates with $I_{t} = \mathrm{In}(t)$ for all $t \in T$, (\ref{eqn:alg_rank_invariant}) implies $[\mathbf{f}_{e'}]_{e' \in \mathrm{In}(t)} = \omega$, and so  $(k_{d,e}(\alpha^{r_j}))$ forms a scalar linear solution over GF($2^{m_L}$) for every $1 \leq j \leq \frac{\phi(L)}{m_L}$. According to Lemma \ref{lemma:scalar_linear_solution_induce_another_scalar}, this in turn guarantees that $(k_{d,e}(\alpha^{r}))$ forms a scalar linear solution for all $r \in C_j$, and as $R = \bigcup_{1 \leq j \leq \frac{\phi(L)}{m_L}} C_j$, every $(k_{d,e}(\alpha^{r}))$, $r \in R$, is a linear solution.

\subsection{List of Notation}
\label{appendix:List_of_Notation}
\begin{longtable}{ll}
$s$:     &     the unique source node.  \\
$T$:     &     the set of receivers.    \\
Out($v$):  &   the set of outgoing edges from node $v$.  \\
In($v$):   &   the set of incoming edges to node $v$.  \\
$\mathbf{m}_{e}$:  & the data unit transmitted on edge $e$, which is an $L$-dimensional row vector. \\
$\omega$: & the number of data units generated by $s$, equal to $|\mathrm{Out}(s)|$ and $|\mathrm{In}(t)|$.   \\
$m_{L}$:  & the multiplicative order of $2$ modulo $L$.  \\
$\alpha$: & the primitive $L^{th}$ root of unity over GF(2), which belongs to GF($2^{m_L}$). \\
$\phi(L)$: & the Euler's totient function of an integer $L$. \\
$\mathbf{I}_{L}$:  & the identity matrix of size $L$. \\
$(\mathbf{K}_{d,e})$:  & the $L$-dimensional linear code over GF($2$), where every local encoding kernel $\mathbf{K}_{d,e}$ \\
& is an $\omega L \times \omega L$ matrix. \\
$\mathbf{F}_{e}$:  & the global encoding kernel for edge $e$, which is an $\omega L \times L$ matrix, of code $(\mathbf{K}_{d,e})$.\\
$(k_{d,e}(\alpha^j))$: & the scalar linear code over GF($2^{m_L}$), where every local encoding kernel $k_{d,e}(\alpha^j)$ is \\
& the evaluation of a defined polynomial $k_{d,e}(x)$ by setting $x$ equal to $\alpha^j$. \\
$\mathbf{f}_e(\alpha)$: & the global encoding kernel for edge $e$ determined by $(k_{d,e}(\alpha))$. \\
$\mathbf{G}_{s}$:  & the $\omega L' \times \omega L$ encoding matrix at source $s$ of an $(L', L)$ linear code. \\
$\mathbf{D}_{t}$:  & the decoding matrix at receiver $t$ of a linear solution.  \\
$\mathbf{C}_{L}$:  & the $L\times L$ cyclic permutation matrix defined in (\ref{eqn:cyclic_permutation_matrix}).\\
$\mathcal{J}$: & the set of integers between $0$ and $L-1$ such that the scalar linear code $(k_{d,e}(\alpha^j))$ \\
& is a linear solution. \\
$J$: & the cardinality of $\mathcal{J}$.
\end{longtable}


\end{document}